\documentclass[12pt,draftcls,onecolumn]{IEEEtran}
\usepackage[normalem]{ulem}
\usepackage[usenames,dvipsnames]{color}
\usepackage{amsfonts}

\usepackage{cite}

 \ifCLASSINFOpdf
   \usepackage[pdftex]{graphicx}
 \else
   \usepackage[dvips]{graphicx}
 \fi

\usepackage[cmex10]{amsmath}

\usepackage{subfig}

\begin{document}
%
\title{Discrete
Imaging Models for Three-Dimensional Optoacoustic Tomography using Radially Symmetric Expansion Functions}
%
%
%

\author{Kun~Wang, ~\IEEEmembership{Member,~IEEE,}
        Robert~W.~Schoonover, 
        Richard~Su,
        Alexander~Oraevsky,~\IEEEmembership{Member,~IEEE,}\\ and
        Mark~A.~Anastasio,~\IEEEmembership{Senior Member,~IEEE}
\thanks{K. Wang, R.W. Schoonover and M.A. Anastasio are with the Department
of Biomedical Engineering, 
Washington University in St. Louis, 
St. Louis, MO 63130
e-mail: wangk@seas.wustl.edu, anastasio@wustl.edu}
\thanks{R. Su and A. Oraevsky are with TomoWave Laboratories,
6550 Mapleridge Street, Suite 124, Houston, TX 77081-4629.}
}

\maketitle

\begin{abstract}
Optoacoustic tomography (OAT), also known as photoacoustic tomography,
is an emerging computed biomedical imaging modality that exploits
optical contrast and ultrasonic detection principles.
Iterative image reconstruction algorithms that are based on
discrete imaging models
are actively being developed  for OAT
due to their ability to improve image
quality by incorporating accurate models
of the imaging physics, instrument response, and measurement noise. 
In this work, we investigate the use of discrete imaging models 
based on Kaiser-Bessel window functions for iterative image reconstruction in OAT.
A closed-form expression for the pressure produced by a Kaiser-Bessel function is calculated, which facilitates accurate computation of the system matrix.
Computer-simulation and experimental studies are employed to demonstrate the potential advantages of
Kaiser-Bessel function-based iterative image reconstruction in OAT.
\end{abstract}


%
\IEEEpeerreviewmaketitle

\section{Introduction}

Optoacoustic tomography (OAT), also referred to as photoacoustic computed
tomography, is an emerging hybrid imaging modality that combines the high
spatial resolution and ability to image relatively deep structures of
 ultrasound imaging with
the high optical contrast of optical imaging \cite{OraBook,WangPABook09}. 
OAT has great potential for use in a
number of biomedical applications, including small animal imaging \cite{WANGX:2003,yang2007functional,petermice:jbo,TumorBeard:2012}, breast imaging \cite{Oraevsky09:Breast,ManoharBC:2010}, and molecular imaging \cite{RazanskiMSOT}.
In OAT, an object is illuminated with short laser pulses that
 result in the subsequent generation of internal acoustic wavefields via the thermoacoustic effect \cite{OraBook,XuReview}.
 The initial amplitudes of the induced acoustic wavefields are proportional to
 the spatially variant absorbed optical energy density within the object, which will be denoted by the object function $A(\mathbf r)$.
 The acoustic wavefields propagate out of the object and are detected by use of a collection of wide-band ultrasonic transducers that are located outside the object. From these acoustic data, an image reconstruction algorithm is employed to obtain an estimate of $A(\mathbf r)$.

As in other tomographic imaging modalities \cite{SparseMRI:2007,XPanIP2009},
 iterative image reconstruction algorithms can improve image quality in
 PACT \cite{AnastasioTATHTfeasible,JCarson:2008,OATTV09:Provost,GuoCS:2010,KunSPIETV:2011,Kun:PMB}. 
Moreover, the development of advanced iterative image reconstruction algorithms can allow for the design of PACT systems that acquire smaller data sets, thus reducing the total data-acquistion time. 
In a previous study, it was demonstrated that iterative image reconstruction algorithms,
 in general,  yield more accurate OAT images than those produced by a mathematically
exact filtered backprojection algorithm  \cite{Kun:PMB}. 
Most OAT iterative reconstruction algorithms are
based on discrete-to-discrete (D-D) imaging models \cite{MarkBookChapter}.
D-D imaging models employ a discrete imaging operator, also known as a system matrix,
  to map  a finite-dimensional approximation of $A(\mathbf r)$ to the measured data vector,
which is inherently finite-dimensional in a digital imaging system.
The finite-dimensional approximation of $A(\mathbf r)$ is often formed as a weighted 
sum of a finite number of expansion functions.
The choice of expansion functions can be motivated by numerous practical and theoretical
considerations that include a desire to minimize representation error, incoporation of  {\em{a priori}}
information regarding the object function, or ease of computation. 
Common choices of expansion functions in OAT include cubic and spherical voxels \cite{GPaltauf:2002,JCarson:2008,TMI:transmodel,Kun:GPU}, and linear
interpolation functions \cite{Jin:2009,Kun:GPU,German3D:12}.  
It should be noted that none of these expansion functions are differentiable at their boundary, and therefore the
pressure signal produced by each of them, when treated as optoacoustic sources, will possess an infinite temporal bandwidth.  As discussed later, this
leads to numerical inaccuracies when computing the associated system matrices.
In general, different choices for the expansion functions will result in system matrices that have
distinct numerical properties \cite{BarrettBook} that will affect the performances of iterative image
reconstruction algorithms.   
There remains an important need for the further development of accurate discrete imaging models for OAT
and an investigation of their ability to mitigate different types of measurement errors found in real-world implementations.

In this work, we develop and investigate a D-D imaging model for OAT based on the use of
 radially symmetric expansion functions known as  Kaiser-Bessel (KB) window functions, also
widely known as `blob' functions in the tomographic reconstruction
 literature \cite{lewitt1990multidimensional,Qiaofeng:12,schweiger2003image}.
Radially symmetric and smooth expansion functions such as these possess a convenient closed-form solution for the optoacoustic pressure signal produced by them, which facilitates accurate OAT system matrix construction. 
KB functions have been
widely employed to establish discrete imaging models for other modalities such as X-ray computed tomography \cite{Lewitt:92,Qiaofeng:12} and optical tomography \cite{schweiger2003image}. 
They have several desirable features that include having
 finite spatial support, being differentiable to arbitrary order at the boundaries, and being quasi-bandlimited.
The statistical and numerical properties of images reconstructed by use of an iterative algorithm
that employs  the KB function-based system matrix
are systematically compared to those corresponding to use of an interpolation-based system matrix.
We also demonstrate the use of non-standard discretization schemes in which the KB  functions are
centered at the verticies of a body centered cubic (BCC) grid rather than a standard 3D Cartesian grid, which
reduces the number of expansion functions required to represent an estimate of  $A(\mathbf r)$ by a factor of $\sqrt{2}$.
It should be noted that the proposed D-D imaging model is general in the sense that the KB functions
can be replaced by any other radially symmetric set of expansion functions that possess a closed-form
solution for the optoacoustic pressure generated by them.
  See, for example, \cite{schweiger2003image}, for  descriptions  of alternative forms of radially symmetric
expansion functions.    

The remainder of the paper is organized as follows. A previously employed linear-interpolation-based
OAT imaging model is reviewed in Section II and the new
KB function-based imaging model is described in Section III.
A description of the numerical and experimental studies are provided in Section IV.
Section V contains the results of these studies
and the paper concludes with a discussion in Section VI.

\section{Background: linear-interpolation-based imaging models}

\subsection{General formulation of discrete-to-discrete (D-D) imaging models}


An  OAT imaging system employing point-like ultrasonic transducers  can be accurately described by a continuous-to-discrete (C-D) imaging model as \cite{MarkBookChapter,TMI:transmodel,Kun:PMB}
\begin{equation}\label{eqn:CDmodel}
  [\mathbf u]_{qK+k} = 
      h^e(t) *_t 
      {\frac{\beta}{4\pi C_p} } 
   \int_{\mathcal V}\!\! d \mathbf{r}\, A(\mathbf{r}) 
    {\frac{d}{dt}} 
    \frac{\delta\left(t-{
         \frac{\vert \mathbf r^s_q-\mathbf{r}\vert}{ c_0}
       }\right)}
  {\vert \mathbf r^s_q-\mathbf r\vert}
      \Bigg|_{t=k\Delta_t},\quad 
  \substack{ q=0,1,\cdots, Q-1\\ 
   k=0,1,\cdots, K-1} , 
\end{equation} 
where $h^e(t)$ is the electrical impulse response (EIR) of the transducer \cite{AConjusteau:2009,TMI:transmodel},
 $*_t$ denotes the temporal convolution operation, $\delta(t)$ is the one-dimensional
 Dirac delta function, and
$\beta$, $c_0$ and $C_p$ denote the thermal coefficient of volume expansion, (constant) speed-of-sound, and the specific heat capacity of the medium at constant pressure, respectively.
The vector  $\mathbf u\in \mathbb{R}^{QK}$ represents a
 lexicographically ordered collection of the sampled values
of the electrical signals that are produced by the ultrasonic transducers employed,
where $Q$ and $K$ denote the  number of transducers employed in the imaging
system and the number of temporal samples recorded
by each transducer, respectively.
The notation $[\mathbf u]_{qK+k}$ will be utilized to denote the 
 $(qK+k)$-th element of $\mathbf u$. Here, the integer-valued indices $q$
and $k$ indicate the transducer position $\mathbf r_q^s\in  \mathbb{R}^{3}$ and temporal sample
acquired with a sampling interval  $\Delta_t$.
The object function $A(\mathbf r)$ is assumed to be bounded
and contained within the volume $\mathcal V$.
 The imaging model can be readily generalized to account
for the spatial impulse reponse of a transducer \cite{TMI:transmodel}.

In practical applications of iterative image reconstruction, it is convenient
to approximate the C-D imaging model in Eqn.\ (\ref{eqn:CDmodel}), which maps the object function
to a finite-dimensional vector, by a fully discrete model.
This requires introduction of
a finite-dimensional representation of $A(\mathbf r)$.
A linear $N$-dimensional approximation of $A(\mathbf r)$, denoted by  $A^a(\mathbf r)$, \cite{BarrettBook, MarkBookChapter} 
can be expressed as
\begin{equation}\label{eqn:dis_obj}
  A(\mathbf r)\approx \sum_{n=0}^{N-1} [\boldsymbol \alpha]_n
    \psi_n(\mathbf r) \equiv A^a(\mathbf r) , 
\end{equation}
where $\boldsymbol\alpha\in \mathbb{R}^N$ is a coefficient vector whose $n$-th component is denoted by $[\boldsymbol\alpha]_n$ and $\{\psi_n(\mathbf r)\}_{n=0}^{N-1}$ is a set of pre-chosen expansion functions. 
On substitution from Eqn.\ \eqref{eqn:dis_obj} into Eqn.\ \eqref{eqn:CDmodel}, one obtains a D-D
 mapping from $\boldsymbol\alpha$ to $\mathbf u$, expressed as
\begin{equation}\label{eqn:DDmodel}
  \mathbf u \approx \mathbf H \boldsymbol \alpha \equiv \mathbf u^a, 
\end{equation}
where the $QK\times N$ matrix $\mathbf H$ is the D-D imaging operator, also known as system matrix, 
whose elements  are defined as 
\begin{equation}\label{eqn:imagoptor}
 [\mathbf H]_{qK+k,n} = 
          h^e(t) *_t   
        {\frac{\beta}{4\pi C_p} } 
   \int_{\mathcal V}\!\! d \mathbf{r}\, \psi_n(\mathbf{r}) 
    {\frac{d}{dt}} 
    \frac{\delta\left(t-{
         \frac{\vert \mathbf r^s_q-\mathbf{r}\vert}{ c_0}
       }\right)}
  {\vert \mathbf r^s_q-\mathbf r\vert}
       \Bigg\vert_{t=k\Delta_t}. 
\end{equation}
The image reconstruction task is to estimate $\boldsymbol \alpha$ by
 approximately inverting Eqn.\ (\ref{eqn:DDmodel}), after which an estimate of
$A(\mathbf r)$ is obtained by use of Eqn.\ (\ref{eqn:dis_obj}).
In principle, the expansion functions $\psi_n(\mathbf r)$ can be arbitrary. 
However, for a given $N$, they should be chosen so that $A(\mathbf r)\approx A^a(\mathbf r)$
and therefore $\mathbf u\approx \mathbf u^a$.  


\subsection{Linear interpolation-based D-D imaging model}

Linear interpolation-based D-D imaging models have been employed for OAT iterative image reconstruction \cite{Jin:2009,German3D:12}. 
These imaging models typically employ spatially-localized expansion functions that  are
centered at the verticies of a Cartesian grid. 
As an example, when a trilinear interpolation method is employed, the expansion function can be expressed as \cite{KakBook,MarkBookChapter}:
\begin{equation}\label{eqn:expfunI}
  \psi^{\rm int}_n(\mathbf r)=\left\{\begin{array}{ll}
                    (1-\frac{|x-x_n|}{\Delta{\rm_s}})
                    (1-\frac{|y-y_n|}{\Delta{\rm_s}})
                    (1-\frac{|z-z_n|}{\Delta{\rm_s}}), & \text{if}\,
  |x-x_n|, |y-y_n|, |z-z_n| \leq  \Delta{\rm_s}\\
                0, & \text{otherwise}
                  \end{array}\right.,
\end{equation}
where $ \mathbf r_n\equiv (x_n,y_n,z_n)$ specifies the location of
 the $n$-th vertex of a Cartesian grid with spacing $\Delta{\rm_s}$. 
For this particular choice of expansion function, the expansion
 coefficient vector will be denoted as $\boldsymbol\alpha_{\rm int}$ and can be 
defined as  $[\boldsymbol \alpha_{\rm{int}}]_n=A(\mathbf r)|_{\mathbf r=\mathbf r_n}$, for $n=0,1,\cdots,N-1$.
The system matrix whoses elements are defined by use of Eqn.\ (\ref{eqn:expfunI}) in Eqn.\ (\ref{eqn:imagoptor})
will be denoted as $\mathbf H_{\rm int}$ and the associated D-D imaging model
is given by
\begin{equation}\label{eqn:ddsysint}
 \mathbf u \approx \mathbf H_{\rm int} \boldsymbol\alpha_{\rm int}.
\end{equation}
Note that the numerical implementation of $\mathbf H_{\rm int}$ requires an additional discretization of the volume integral in Eqn.\ \eqref{eqn:imagoptor}.  
Details regarding the numerical implementation of  $\mathbf H_{\rm int}$ can be found in Ref.\ \cite{Kun:GPU}.

\section {Kaiser-Bessel function-based OAT imaging models}
Below we establish a D-D imaging model for OAT that is based on the
use of KB expansion functions.
The imaging model will incorporate both the electrical and spatial impulse responses of the ultrasonic transducers employed. 

\subsection{Kaiser-Bessel expansion functions in OAT}

The KB function of order $m$ is defined as \cite{lewitt1990multidimensional,schweiger2003image} 
\begin{eqnarray}\label{eqn:KBbasis}
b(x) = \left\{ \begin{array}{l l} \left(\sqrt{1-x^2/a^2}\right)^m \frac{I_m\left(\gamma\sqrt{1-x^2/a^2}\right)}{I_m(\gamma)} & 0\leq x \leq a \\ 0 & a<x ,\end{array}\right.
\end{eqnarray}
where $x\in \mathbb R^+$, $I_m(x)$   is the modified Bessel function of the first kind of
 order $m$, and $a\in \mathbb R^+$ and $\gamma\in \mathbb R^+$ determine the support
 radius and the smoothness of $b(x)$, respectively. 
Following previously employed terminology \cite{Lewitt:92}, we refer to the expansion
 function $\psi^{\rm KB}_n(\mathbf r) \equiv b(x)|_{x=|\mathbf r - \mathbf r_n|}$ as a KB  function
centered at location $\mathbf r_n$.

The system matrix whose elements are defined by  use of $\psi^{\rm KB}_n(\mathbf r)$ in
Eqn.\ (\ref{eqn:imagoptor}) will be denoted by $\mathbf H_{\rm KB}$.
Unlike with $\mathbf H_{\rm int}$, the elements of
 $\mathbf H_{\rm KB}$ can be computed analytically, as described below.
This is highly desirable, as it circumvents the need to numerically approximate
Eqn.\ (\ref{eqn:imagoptor})  \cite{Bob13:blob}.
In contrast, the linear interpolation-based models usually require 
numerical approximations to compute the system matrix \cite{Kun:GPU}, which can introduce errors
that ultimately degrade the accuracy of the reconstructed image.
A similar phenomenon has been analyzed in differential X-ray phase-contrast tomography 
 image reconstruction \cite{Qiaofeng:12}. 
Several linear interpolation methods have been proposed to analytically calculate the imaging operator acting on each voxel,
but numerical instabilities are present corresponding to certain tomographic view angles \cite{Ntziachristos:QPAT2012}.

It will prove convenient to formulate the KB function-based imaging model in
the temporal frequency domain \cite{Kun:PMB}.
Consider that the discrete Fourier transform (DFT) of the sampled  temporal data recorded
by each transducer is computed.
Let $ \tilde {\mathbf u}$ denote a temporally Fourier transformed data
vector formed by lexicographically ordering these data.
 The imaging model in the temporal-frequency domain will be expressed as
\begin{equation}\label{eqn:ddsysblob}
  \tilde {\mathbf u} \approx \tilde {\mathbf H}_{\rm KB} \boldsymbol\alpha_{\rm KB}. 
\end{equation}
The elements of the modified system matrix $\tilde{\mathbf H}_{\rm KB}$ are given by \cite{Kun:PMB}
\begin{equation}\label{eqn:blobsysmatrix}
  [\tilde{\mathbf H}_{\rm KB}]_{qL+l,n} = \tilde{p}_0^{\rm KB}(f)
       \tilde{h}^e(f)
       \tilde{h}_q^s(\mathbf r_n,f)\big|_{f=l\Delta_f},\,\, {\rm for}\,\, l = 0,1,\cdots,L-1,  
\end{equation}
where $\Delta_f$ denotes the temporal frequency sampling interval, $\tilde h^e(f)$ is the one-dimensional
 Fourier transform of $h^e(t)$,
 and $\tilde h _q^s (\mathbf r_n, f)$ is the spatial impulse response (SIR) in the temporal frequency domain \cite{TMI:transmodel,Kun:PMB,JBO13:cylindrical}. 
When a point-like transducer assumption is justified, $\tilde h _q^s (\mathbf r_n, f)$ degenerates to the Green function 
\begin{equation}
  \tilde h _q^s (\mathbf r_n, f) = \frac{\exp(-\hat\jmath 2\pi f \frac{ |\mathbf r^s_q-\mathbf r_n|}{c_0})}{2\pi |\mathbf r^s_q-\mathbf r_n|},
\end{equation}
where $\mathbf r_q^s$ and $\mathbf r_n$ are locations of the $q$-th transducer and the center of the $n$-th KB
function, respectively. 
The quantity $\tilde p_0^{\rm KB}(f)$ is the temporal Fourier transform of the acoustic pressure generated by a KB
function located at the origin and is expressed as \cite{diebold2009photoacoustic,Bob13:blob} (See Appendix) 
\begin{equation}\label{eqn:p0blob}
  \tilde p_0^{\rm KB}(f) = - \frac{\hat\jmath 4 \pi^2 f a^3 \beta}{C_p I_m(\gamma)} \frac{j_{m+1}(\sqrt{4\pi^2a^2f^2/c_0^2-\gamma^2})}{(4\pi^2a^2f^2/c_0^2-\gamma^2)^{(m+1)/2}},
\end{equation}
where $j_m(x)$ is the $m$-th order spherical Bessel function of the first kind. Equation~(\ref{eqn:blobsysmatrix}) is valid for any radially symmetric expansion function.  
Note that a previously proposed OAT imaging model that employed
 uniform spherical voxels as the expansion functions
 \cite{TMI:transmodel,Kun:PMB} is contained
as a special case of the KB function-based imaging model corresponding to $m=0$, $\gamma=0$, and $a=\Delta{\rm_s}/2$. 

{Selection of parameters for the KB  function in Eqn.~\eqref{eqn:KBbasis} has been comprehensively described in the literature \cite{matej1996practical}.  
The parameter $m$, for example, determines the differentiability of the expansion function, $b(x)$ at $x=a$. 
In applications in which the derivative of the expansion function appears in the imaging model, $m \geq 2$ is chosen so that the derivative is continuous at the KB function boundary. 
The choice of the parameter $a$, which determines the effective voxel size, is determined by the size of the reconstruction volume and the desired resolution.  
In general, $a$ is chosen to be comparable to the size of the finest feature of interest,
otherwise an overshoot may be observed in the reconstructed images. 
However, reducing the value of $a$ will lead to an increase of computational demands. 
The parameter $\gamma$ affects the bandwidth of the individual expansion elements.  
In Fig.~\ref{blob_profiles}, normalized plots of Eq.~(\ref{eqn:p0blob}) are shown for four values of $\gamma$: $\gamma = 1, 4, 7$, and $10.4$, with $m = 2$ and $a = 0.28$ mm.  One sees immediately that the bandwidth of the the KB function increases monotonically with increasing $\gamma$.  A similar effect can be achieved by decreasing the parameter $a$ while keeping $m$ and $\gamma$ fixed.
It may be reasonable in some circumstances to tune the value of $\gamma$ to the measured bandwidth of the measured pressure signal.  
The choice of $\gamma = 10.4$ is often made in both X-Ray CT \cite{lewitt1990multidimensional}, and optical tomography \cite{schweiger2003image} because it provides the smallest representation error when estimating a piecewise constant function \cite[see, for example, Fig. 4]{schweiger2003image}. 
The choice of optimal parameters is, however, application-dependent
\cite{furuie1994methodology,matej1994evaluation}. 
}

\subsection{Kaiser-Bessel functions on non-standard grids} 
\label{sect:bcc}

The expansion functions $\{\psi_n(\mathbf r)\}$ are typically positioned on a 3D
 Cartesian grid  when constructing  D-D imaging models for OAT, including the linear-interpolation-based imaging models. 
The Cartesian grid,  also referred to as simple cubic (SC) grid, is a natural
 choice if the support volume of $\psi_n(\mathbf r)$ is cubic.
When the support volume is a sphere, however, 
  body centered cubic (BCC) and face centered (FCC) grids, as sketched in Fig.\ \ref{fig:gridstructure}, can have advantages and 
have been proposed for use in X-ray computed tomography\cite{Lewitt95:BCC}. 
Let $\Delta{\rm_s}$, $\Delta{\rm_s^b}$, and $\Delta{\rm_s^f}$ denote the grid spacing of the 
SC, BCC and FCC grids, respectively. 
When the grid spacing satisfies $\Delta{\rm_s^b}=\sqrt{2} \Delta{\rm_s}$ and $\Delta{\rm_s^f}=\sqrt{3} \Delta{\rm_s}$, the three types of grids will be referred to as 
``equivalent'' \cite{Lewitt95:BCC} because the highest spatial-frequency of the object function is equivalently limited by $1/(2\Delta{\rm_s})$ if an unaliased sampling is desired.
Accordingly, the BCC and the FCC grids can potentially reduce the number
 of required expansion functions by factors of $\sqrt 2$ and $3\sqrt{3}/4$ respectively \cite{Lewitt95:BCC}. 
Unlike with an FCC grid, the implementation of an imaging model corresponding to
 a BCC grid is very similar to the implementation of one corresponding to a
 SC grid because the BCC grid can be interpreted as two interleaved SC grids. 
In the numerical studies described below, we investigate the use of the KB function-based imaging model for 3D OAT assuming a BCC grid with spacing $\Delta{\rm_s^b}=\sqrt{2}\Delta{\rm_s}$.



\section{Descriptions of Numerical Studies }

Numerical studies were conducted to compare 
 the numerical properties of the system matrices $\mathbf H_{\rm int}$ and $\tilde{\mathbf H}_{\rm KB}$
and analyze differences in the numerical and statistical properties of images reconstructed by use of them.

\subsection{Simulation of noise-free data and imaging geometry}

In this work, the numerical phantoms representing $A(\mathbf r)$ consisted of a collection of spheres.
 Each sphere possessed a different center location, radius and absorbed optical energy density, denoted
 by $\mathbf r_i$, $R_i$ and $A_i$ for the $i$-th sphere.  The noise-free data  for the phantoms were simulated by two steps:
first, samples of the acoustic pressure generated by each spherical structure were analytically calculated as \cite{WangPABook09,MarkBookChapter} 
\begin{equation}
  p_i (\mathbf r^s_q, t)|_{t=k\Delta_t} =  
    \left\lbrace\begin{array}{ll}
       A_i\Big[- \frac{\beta c_0^3}{C_p |\mathbf r^s_q - \mathbf r_i|} t
        + \frac{\beta c_0^2}{2} \Big]_{t=k\Delta_t},
         & {\rm if} \; \big|c_0k\Delta_t-|\mathbf r^s_q -\mathbf r_i| \big| \leq R_i \\
        0, & {\rm otherwise.} \end{array}\right. 
\end{equation}
Second, the resulting $p_i (\mathbf r^s_q, t)|_{t=k\Delta_t}$ were subsequently convolved with $h^e(t)$  and  summed to generate the noise-free data as 
\begin{equation} \label{eqn:noiseFree}
  [\mathbf u]_{qK+k}= h^e(t) *_t \sum_{i=0}^8 
             p_i (\mathbf r^s_q, t) \big\vert_{t=k\Delta_t},
\end{equation}
{where $h^e(t)$ was experimentally measured  \cite{AConjusteau:2009}
(3 MHz bandwidth with 3 MHz center frequency.)
We ignored the SIR in order to facilitate the implementation of the linear-interpolation-based imaging model. 
Also the point-like transducer assumption is likely to be sufficiently accurate for our experimental system when the object is located near the center \cite{petermice:jbo,TMI:transmodel,Kenji14:farfield}.  
From the time domain data $\mathbf u$, the temporal-frequency domain data $\tilde{\mathbf u}$ were computed by use of the  fast Fourier transform (FFT) algorithm.

The simulated imaging system is described as follows. 
We employed a spherical measurement surface of radius $R^s=65$ mm centered at the origin of a global coordinate system as shown in Fig.\ \ref{fig:geo}-(a). 
The measurement surface was divided by $N_r=48$ circles of latitude and $N_v=96$ semi-circular arcs of longitude of equiangular intervals in both polar and azimuth angles. 
The intersections of these circles and arcs define the locations of $4608$ point-like transducers. 
The sampling rate was $20$ MHz.
The dimension of the measurement surface is consistent with the experimental system described in Section \ref{sec:exp}. 
Each transducer acquired $N_t=256$ time samples, or equivalently $N_f=256$ temporal-frequency samples computed by use of the FFT algorithm. 
The object was contained in a cube of size $8.96$ mm in each dimension that was centered at the origin.

\subsection{Image reconstruction algorithms}
\label{sect:reconalgorithm}

Image reconstruction was conducted by first solving 
\begin{equation}\label{eqn:costfuncint}
  \hat{\boldsymbol\alpha}_{\rm int}=\arg\min_{\boldsymbol\alpha} 
          \Vert \mathbf u - \mathbf H_{\rm int} \boldsymbol\alpha \Vert^2
             +  \beta_{\rm int} R (\boldsymbol\alpha) ,
\end{equation}
and 
\begin{equation}\label{eqn:costfuncbcc}
  \hat{\boldsymbol\alpha}_{\rm KB}=\arg\min_{\boldsymbol\alpha} 
          \Vert \tilde{\mathbf u} - \tilde{\mathbf H}_{\rm KB} \boldsymbol\alpha \Vert^2
             +  \beta_{\rm KB} R (\boldsymbol\alpha),
\end{equation}
to estimate the expansion coefficients for the linear-interpolation- and KB function-based imaging models respectively. 
Here, $R(\boldsymbol\alpha)$ is the regularization penalty and $\beta_{\rm int}$ and $\beta_{\rm KB}$ are regularization parameters. 
A conventional quadratic penalty was employed to promote local smoothness, i.e.,  
\begin{equation}\label{eqn:penalty}
  R(\boldsymbol\alpha) = \sum_n \sum_{i\,\in\,\mathcal N(n)} ([\boldsymbol\alpha]_n-[\boldsymbol\alpha]_i)^2, 
\end{equation}
where $\mathcal N(n)$ is an index set of the neighboring voxels of the $n$-th voxel.
We implemented a linear conjugate gradient algorithm to solve Eqns.\ \eqref{eqn:costfuncint} and \eqref{eqn:costfuncbcc} iteratively based on the associated normal equations \cite{Shewchuk:1994}. 
The iteration was terminated when the residual of the cost function was reduced to a prechosen level in its Euclidean norm \cite{Shewchuk:1994}. 
From the resulting coefficient vectors $\hat {\boldsymbol \alpha}_{\rm int}$ and $\hat{\boldsymbol\alpha}_{\rm KB}$, images were estimated by use of Eqn.\ \eqref{eqn:dis_obj}, rewritten as 
\begin{equation}\label{eqn:intA}
  \hat A_{\rm int} (\mathbf r) = \sum_{n=0}^{N_{\rm int}-1} [\hat{\boldsymbol\alpha}_{\rm int}]_n \psi_n^{\rm int}(\mathbf r)
\end{equation}
and 
\begin{equation}\label{eqn:blobA}
  \hat A_{\rm KB} (\mathbf r) = \sum_{n=0}^{N_{\rm KB}-1} [\hat{\boldsymbol\alpha}_{\rm KB}]_n \psi_n^{\rm KB}(\mathbf r),
\end{equation}
for the linear-interpolation- and KB function-based imaging models respectively, where $N_{\rm int}$ and $N_{\rm KB}$ are the total number of corresponding expansion functions.


\subsection{Singular value analysis of D-D imaging models}
\label{subsect:svd}

A singular value analysis was conducted to gain insights into the intrinsic stability of image reconstruction by use of system matrices $\mathbf H_{\rm int}$ and $\tilde{\mathbf H}_{\rm KB}$.
We reduced the number of rows of both  $\mathbf H_{\rm int}$ and $\tilde{\mathbf H}_{\rm KB}$ to circumvent the great demand of memory in the calculation of singular values. 
More specifically, if the reduced-dimensional system matrices $\mathbf H_{\rm int}$ (or $\tilde{\mathbf H}_{\rm KB}$) act on $\boldsymbol\alpha_{\rm int}$ (or $\boldsymbol\alpha_{\rm KB}$), the resulting vector will estimate the voltage signals (or the temporal-frequency spectra) received by a single transducer located at $(R^s,0,0)$ mm. 
We expect the singular value spectra of the reduced-dimensional system matrices to be similar to those of the original system matrices because the imaging system is approximately rotationally symmetric.
The relation between the singular values of the reduced system matrices and those of the original system matrices can be found in \cite{Kupinski10:SVD}. 
The QR and QZ algorithms \cite{smith1976:eispack} embedded in MATLAB were employed to calculate the eigenvalues of the reduced-dimensional $\mathbf H_{\rm int}\mathbf H^\dagger_{\rm int}$ and $\tilde{\mathbf H}_{\rm KB}\tilde{\mathbf H}_{\rm KB}^\dagger$ respectively. 
By taking the square root of the eigenvalues, singular value spectra of the reduced-dimensional $\mathbf H_{\rm int}$ and $\tilde{\mathbf H}_{\rm KB}$ were obtained.

\subsection{Simulation of random object functions}

In order to investigate the effect of representation errors on the reconstructed images, we employed a random process to generate an ensemble of object functions \cite{furuie1994methodology}. 
The random object function will be denoted by $\underline{A}(\mathbf r)$.
Here and throughout this manuscript, the underline indicates that the corresponding quantity is random. 
Each realization of $\underline{A}(\mathbf r)$ consisted of $9$ smooth spheres (indexed by $i$ for $i=0,1,\cdots, 8$) with random center locations, radii, and absorbed optical energy densities, denoted by $(\underline{x}_i,\underline{y}_i,\underline{z}_i)$, $\underline{R}_i$, and $\underline{A}_i$, respectively. 
A slice through the plane $z=0$ of a single realization of $\underline{A}(\mathbf r)$ is provided in Fig.\ \ref{fig:geo}-(b).  
The statistics of $\underline{A}(\mathbf r)$ are listed in Table \ref{tab:phantom}, where the standard deviations (STD) are given in units of either mm or percentage of the corresponding mean values. 
The spheres indexed from $1$ to $5$ were blurred by use of Gaussian kernels $G_i(\mathbf r)$ whose full width at half maximums (FWHM) are also given in Table \ref{tab:phantom}. The blurring of the spheres was implemented by modifying Eqn.~(\ref{eqn:noiseFree}) as 
\begin{equation}
  [\mathbf u]_{qK+k}= h^e(t) *_t \sum_{i=0}^8 
             p_i (\mathbf r^s_q, t)  *_t g_i(t) \big\vert_{t=k\Delta_t},
\end{equation}
where $g_i(t)$ is a Gaussian kernel whose FWHM is that of $G_i(\mathbf r)$ scaled by a factor of $1/c_0$ \cite{FourierShell:07}.
We generated $64$ realizations of $\underline{A}(\mathbf r)$, each of which will be denoted by $A^{(j)}(\mathbf r)$ for $j=0,1,\cdots,63$. 


\subsection{Simulation of measurement noise}

In order to analyze the noise properties of $\mathbf H_{\rm int}$ and $\tilde{\mathbf H}_{\rm KB}$, an additive Gaussian white noise model was employed to simulate electronic noise:
\begin{equation}
  \underline{\mathbf u} = \mathbf u + \underline{\mathbf n}, 
\end{equation}
where $\underline{\mathbf n}$ is the Gaussian white noise process, $\mathbf u$ is the noiseless voltage data corresponding to $A(\mathbf r)$, and $\underline{\mathbf u}$ is the measured noisy data. 
The STD of $\underline{\mathbf n}$ was set to be $10\%$ of the maximum of $\mathbf u$.
We simulated $128$ realizations of $\underline{\mathbf u}$. 
The corresponding temporal-frequency domain data $\tilde{\underline{\mathbf u}}$ were computed by use of the FFT algorithm.

\subsection{Assessment of reconstructed images}
\label{sect:re-sample}

The accuracy of a reconstructed image, in principle, can be assessed by an error functional \cite{BarrettBook}
\begin{equation}\label{eqn:reperrorC}
\mathcal E (\hat A) =  \Vert A(\mathbf r)-\hat A(\mathbf r)\Vert^2
\equiv \int_V\!\!d\mathbf r\, [A(\mathbf r)- \hat A(\mathbf r) ]^2, 
\end{equation}
where $\hat A(\mathbf r)$ is the finite-dimensional representation of $A(\mathbf r)$ that is specified by the estimated coefficient vector, and $\mathcal E(\hat A)$ measures the squared Euclidean distance from $\hat A(\mathbf r)$ to $A(\mathbf r)$. 
Because the volume integral in Eqn.~\eqref{eqn:reperrorC} lacks a closed-form solution, a numerical approximation was employed as  
\begin{equation}\label{eqn:reperrorA}
  \mathcal E (\hat A) \approx  \Delta_{\rm d}^3 \sum_{m=0}^{M-1} [ A(\mathbf r_m) - \hat A_d (\mathbf r_m)]^2,
\end{equation}
where $\hat A_d (\mathbf r_m)$ denotes the estimation of the object function found by sampling $\hat{A}({\bf{r}})$ onto a fine SC grid and $\mathbf r_m$ specifies the location of the $m$-th vertex on the fine SC grid with spacing $\Delta_{\rm d}$. 
The grid spacing $\Delta_{\rm d}$ is required to be smaller than $\Delta_{\rm s}$ to justify the approximation in Eqn.~\eqref{eqn:reperrorA}. 
The fine SC grid will be referred to as a ``display grid'' and is used throughout the manuscript to compare reconstructions using the linear interpolation- and KB function-based image reconstruction algorithms. 
Furthermore, in order to investigate the dependence of reconstruction accuracy on various object structural features, regional mean-square errors (MSE) are introduced as 
\begin{equation}\label{eqn:MSE}
 {\rm MSE} = \frac{1}{M_r} \sum_{m\in\mathcal S_r} [ A(\mathbf r_m) - \hat A_d(\mathbf r_m) ]^2,  
\end{equation}
where $\mathcal S_r$ is the index set of display grid vertices contained within a certain ROI, 
and $M_r$ is the dimension of $\mathcal S_r$. 
We defined $5$ ROIs (see Fig.~\ref{fig:noiselessZ}(a)) that contain different features of the numerical phantom, including a sharp small structure (box 0), a sharp edge (box 1), a moderately blurred edge (box 2), a slowly varying region (box 3) and a uniform region (box 4). 
Note that all the ROIs are $3$D volumes of dimension $0.56^3$ mm$^3$ and their locations are associated with the structures, which vary among the realizations of $\underline{A}(\mathbf r)$.
For the object function  ${A}^{(0)}(\mathbf r)$, all ROIs are centered in the plane $z=0$ and are marked in Fig.~\ref{fig:noiselessZ}-(a).  
Besides the $3$D ROIs, we also calculated the regional MSE across the $2$D plane $z=0$ as an overall accuracy measure. 
For both the 3D ROIs and the 2D plane z=0, the MSE  was calculated for each realization of
the object function.  Due to object variablity, the MSE for each realization of the object function is
random and will be denoted by $\underline{{\rm MSE}}$.
From the ensemble of object functions, the ensemble mean-square error (EMSE) was calculated as 
\begin{equation}
  {\rm EMSE} = \frac{1}{J} \sum_{j=0}^{J-1} {\rm MSE}^{(j)},
\end{equation}
where the $ {\rm MSE}^{(j)}$ denotes the $j$-th realization of $\underline{{\rm MSE}}$.

The accuracy of reconstructed noisy images were quantified by their first- and second-order statistics.  
From $J$ noisy realizations, the mean and variance of reconstructed images were estimated by 
\begin{equation}
  {\rm Mean}_{\rm A} (\mathbf r) \approx \frac{1}{J}\sum_{j=0}^{J-1} \hat{A}^{(j)}(\mathbf r), 
\end{equation}
and 
\begin{equation}
  {\rm Var}_{\rm A} (\mathbf r) \approx
      \frac{1}{J-1}\sum_{j=0}^{J-1} ( \hat{A}^{(j)}(\mathbf r) - {\rm Mean}_{\rm A} (\mathbf r) )^2,
\end{equation}
respectively. 
Because the statistics of the reconstructed images depend on the regularization parameter \cite{Kun:PMB,Fessler:blob94,qi12:4DSPECT}, we swept the regularization parameter over a wide range to generate a curve of ${\rm Var}_{\rm A}$ against ${\rm Mean}_{\rm A}$ for each system matrix. 
From these curves, we investigated the performance of $\mathbf H_{\rm int}$ and $\tilde{\mathbf H}_{\rm KB}$ on balancing the bias and variance of the reconstructed images.

\subsection{Experimental validation}
\label{sec:exp}

We investigated the performance of $\mathbf H_{\rm int}$ and $\tilde{\mathbf H}_{\rm KB}$ by use of experimentally measured data. 
The experimental data were collected by use of a custom-built optoacoustic imaging module \cite{petermice:jbo,Kun:PMB}.
The ultrasonic transducer array (Imasonic SAS, Voray sur l'Ognon, France) contained $64$ piezo-composite ultrasound transducers ($1.5$-$4.5$ MHz bandpass at $-6$ dB) uniformly mounted on an arc-shaped array of radius $65$ mm and subtended angle $152^\circ$.
Targets were positioned in the center and rotated by a stepper motor (DGM60-ASAK Oriental Motor, Tokyo, Japan).
The targets were encased in a water tank that had a pump (Rena FilStar XP1, Surrey UK), PID controller (Auber SYL 1512A, Alpharetta, GA) and heater (Hydor ETH300, Sacramento, CA) in order to maintain a controlled water temperature of $27.1 ^\circ$ C. 
Two randomized bifurcated fiber bundles were oriented orthogonally with respect to the probe that had an output profile of $1$ mm by $50$ mm coming from outside the water tank. 
These fibers were attached to a tunable Q-switched laser system (SpectraWave, TomoWave Laboratories, Houston, TX) operating at $10$ Hz with output wavelength of $780$ nm. 
Data acquisition was performed with analog amplifiers set to $75$ dB with a sampling rate of 20 MHz. 
More details regarding the system can be found in \cite{petermice:jbo,Rich12:JBO}. 

A phantom was built that contained transparent $10\%$ gelatin shaped in a cylinder of radius $25.4$ mm and height $100$ mm, as shown in Fig.~\ref{fig:target}.  
Embedded in the phantom were two plastisol spheres of $7$ mm diameter.  
The right sphere shown in Fig.~\ref{fig:target} possessed a larger absorbing coefficient at the illumination wavelength of $780$ nm. 
Additionally, on one end of the cylinder an acrylic hollow cylinder was embedded about $15$ mm deep in order to attach the phantom to the rotational motor.
During the scanning, both the phantom and the transducer array were oriented vertically, i.e., parallel to the z-axis in the coordinate system shown in Fig.~\ref{fig:geo}. 
The transducer array was fixed while the phantom was rotated about the z-axis over $360^\circ$ with a step size of $2.4^\circ$, resulting in a partially covered spherical measurement surface. 
At each transducer location, $1024$ temporal samples were acquired for two consecutive illuminations and then averaged together, improving the signal-to-noise ratio. 
Accordingly, the dimension of the measured data set was $1024\times150\times63$.
Note that the data acquired by the first element on the 64-element transducer array were employed for
time alignment intead of for image reconstruction. 
We repeated the data acquisition procedure described above $64$ times, creating an ensemble of noisy measurements. 


Images were reconstructed by first solving the penalized least-squares objectives defined by Eqns.~\eqref{eqn:costfuncint}, \eqref{eqn:costfuncbcc} and \eqref{eqn:penalty}, where the system matrices $\mathbf H_{\rm int}$ and $\tilde{\mathbf H}_{\rm KB}$ were calculated on the fly \cite{Kun:GPU}. 
The phantom was contained in a volume of dimension $14.0\times14.0\times32.2$ mm$^3$. 
For the reconstructions, the expansion functions were chosen to be $\{\psi_n^{\rm int}(\mathbf r)\}_{n=0}^{N_{\rm int}-1}$ distributed on a SC grid of spacing $\Delta_{\rm s} = 0.56$ mm and $\{\psi_n^{\rm KB}(\mathbf r)\}_{n=0}^{N_{\rm KB}-1}$ distributed on a BCC grid of spacing $\Delta_{\rm s}^{\rm b} = 0.8$ mm, respectively. 
For the KB function-based imaging model, we let $a=2\Delta{\rm_s}$, $\gamma=10.4$, and $m=2$ in Eqn.\ \eqref{eqn:KBbasis} \cite{Lewitt95:BCC}.  
Accordingly, $\mathbf H_{\rm int}$ and $\tilde{\mathbf H}_{\rm KB}$ were of dimension $(63\times150\times1024)$-by-$(25\times25\times58)$ and $(63\times150\times1024)$-by-$(18\times18\times40\times2)$, respectively (thus $N_{\rm int} = 36,250$ and $N_{\rm KB} = 25,920$).
The values of $N_{\rm int}$ and $N_{\rm KB}$ were chosen so that the size of the reconstructed volume approximately matched the size of the original experimental volume.
From the estimated coefficient vectors $\hat{\boldsymbol\alpha}_{\rm int}$ and $\hat{\boldsymbol\alpha}_{\rm KB}$, $\hat A_{\rm int}(\mathbf r)$ and $\hat A_{\rm KB}(\mathbf r)$ were determined by use of Eqns.~\eqref{eqn:intA} and \eqref{eqn:blobA}.

Image quality was assessed based on a parameter-estimation task. 
The parameter to be estimated was the average value within an ROI of size $1\times 1$ mm$^2$ in a single plane of the object, denoted by $\theta_{\rm true}$.  
We set $\theta_{\rm true}$ to be the one estimated from a reference image as 
\begin{equation}\label{eqn:reftheta}
  \theta_{\rm true} = \frac{1}{M_r}\sum_{ m\in \mathcal S_r} {\hat A}^{\rm ref}_d(\mathbf r_m), 
\end{equation}
where  ${\hat A}^{\rm ref}_d(\mathbf r_m)$ denotes the reference image, evaluated at $\mathbf r_m$, that was iteratively reconstructed by use of $\mathbf H_{\rm int}$ with $\Delta{\rm_s}=0.14$ mm and $\beta_{\rm int} = 1\times10^{-2}$ from the data averaged over the $64$ noisy measurements.
Estimates of $\theta_{\rm true}$ from noisy measurements, denoted by $\underline\theta$, were calculated by  
\begin{equation}
  \underline \theta = \frac{1}{M_r}\sum_{ m\in \mathcal S_r} \underline {\hat A}_d(\mathbf r_m), 
\end{equation} 
where $\underline {\hat A}_d(\mathbf r_m)$ is the random image, evaluated at $\mathbf r_m$. 
We employed the bias and variance of $\underline\theta$ as the figures of merit to evaluate the quality of images reconstructed by use of $\mathbf H_{\rm int}$ and $\tilde{\mathbf H}_{\rm KB}$.
The bias of $\underline\theta$ was estimated by  
\begin{equation}\label{eqn:biassim}
  {\rm Bias}_{\rm \theta} \approx \frac{| \frac{1}{J}\sum_{j=0}^{J-1} \theta^{(j)}  - \theta_{\rm true} |}{\theta_{\rm true}}
    \times 100\%, 
\end{equation}
where $J$ is the number of realizations of $\underline\theta$, 
Note that this choice of reference in Eqn.~\eqref{eqn:reftheta} actually favors the performance of $\mathbf H_{\rm int}$. 
Also, the variance of $\underline\theta$ was estimated by  
\begin{equation}\label{eqn:varsim}
  {\rm Var}_{\rm \theta} \approx \frac{1}{J-1} \sum_{j=0}^{J-1} \Big(
      \theta^{(j)} - 
   \frac{1}{J} \sum_{j'=0}^{J-1} \theta^{(j')} 
    \Big)^2.
\end{equation}
We swept the regularization parameter over a wide range to investigate the performance of $\mathbf H_{\rm int}$ and $\tilde{\mathbf H}_{\rm KB}$ on balancing the tradeoff between $\rm Bias_\theta$ and $\rm Var_\theta$ \cite{Kun:PMB,Fessler:blob94,qi12:4DSPECT}. 

\section{Numerical results}

\subsection{Singular value analysis of the D-D imaging models}
\label{sect:svdresults}

Singular value spectra of $\mathbf H_{\rm int}$ and $\tilde{\mathbf H}_{\rm KB}$ were calculated with equivalent SC and BCC grids, respectively, i.e, $\Delta{\rm_s^b}=\sqrt{2}\Delta{\rm_s}$. 
Two grid spacing values were investigated for both $\mathbf H_{\rm int}$ and $\tilde{\mathbf H}_{\rm KB}$ respectively.
For $\Delta{\rm_s}=0.07$ and $0.14$ mm, $\mathbf H_{\rm int}$ was of dimension $256$-by-$128^3$ and $256$-by-$64^3$ respectively. 
For $\Delta{\rm_s^b}=0.1$ and $0.2$ mm, $\tilde{\mathbf H}_{\rm KB}$ was of dimension $256$-by-$(90^3\times2)$ and $256$-by-$(45^3\times2)$ respectively.
We set $a=\sqrt{2}\Delta{\rm_s^b}$, $\gamma=10.4$, and $m=2$ in Eqn.\ \eqref{eqn:KBbasis} for the calculation of $\tilde{\mathbf H}_{\rm KB}$. 
These values were chosen to minimize the number of expansion elements while limiting representation errors \cite{Lewitt95:BCC}. 

The singular value spectra of $\tilde{\mathbf H}_{\rm KB}$ is, in general, spread over a wider range compared to that of  $\mathbf H_{\rm int}$ as shown in Fig.~\ref{fig:svd}. 
Note that only the first $\sim$$160$ singular values of $\mathbf H_{\rm int}$ fall above our truncation threshold of $10^{-4}$.  
Since both $\mathbf H_{\rm int}\mathbf H_{\rm int}^\dagger$ and $\tilde{\mathbf H}_{\rm KB}\tilde{\mathbf H}_{\rm KB}^\dagger$ are of dimension $256$-by-$256$, the results suggest that the condition number of $\tilde{\mathbf H}_{\rm KB}$ is smaller than that of $\mathbf H_{\rm int}$.
Therefore, use of $\tilde{\mathbf H}_{\rm KB}$ should, in principle, result in a faster convergence rate for certain gradient-based optimization algorithms, including the conjugate gradient algorithm \cite{CGconverge:04}. 
However, the faster convergence rate is of limited practical interest because the iteration is almost always terminated before the final convergence is achieved. 
If measurement noise can be  approximated as white, the singular value spectra also suggest that iterative image reconstruction based on $\tilde {\mathbf H}_{\rm KB}$ is more robust to measurement noise because the singular values of $\tilde {\mathbf H}_{\rm KB}$ are in general larger than those of $\mathbf H_{\rm int}$ \cite{BarrettBook}. 
Note that when using the reduced grid spacing, the singular values have larger magnitudes than with the coarser spacing in the range of the $70$-th to $130$-th singular value, suggesting more components of the object function can be stably reconstructed. 
This gain, however, is traded with a cubical increase in computational time. 

\subsection{Images reconstructed from an ensemble of noiseless data}
\label{sect:noiseless}

Images were reconstructed from noiseless simulated measurement data by use of a least-squares (LS) objective, i.e.  $\beta_{\rm int} = \beta_{\rm KB}=0$ in Eqns.\ \eqref{eqn:costfuncint} and \eqref{eqn:costfuncbcc}. 
We set $\Delta{\rm_s} = 0.14$ mm, $\Delta{\rm_s^b} = 0.2$ mm, $a=0.28$ mm, $\gamma=10.4$, and $m=2$.  
Accordingly, $\hat{\boldsymbol\alpha}_{\rm int}$ and $\hat{\boldsymbol\alpha}_{\rm KB}$ were of dimensions $64^3$ and $45^3\times2$, respectively. 
In addition, a display grid of spacing $\Delta{\rm_d}=0.0175$ mm was selected for image quality assessment as described in Sect.~\ref{sect:re-sample}. 

Images reconstructed by use of $\tilde{\mathbf H}_{\rm KB}$, shown in Fig.~\ref{fig:noiselessZ}-(c), are more accurate than those reconstructed by use of $\mathbf H_{\rm int}$ as shown in Fig.~\ref{fig:noiselessZ}-(b). 
The $\rm MSE$ of the 2D slice in the plane of $z=0$ of the image reconstructed by use of $\tilde{\mathbf H}_{\rm KB}$ (${\rm MSE}=3.50\times10^{-3}$) is only $13.3\%$ of that by use of $\mathbf H_{\rm int}$ (${\rm MSE}=26.32\times10^{-3}$). 
Here, iterations were terminated when the Euclidean norm of the residual of the cost functions was reduced to $0.01\%$  
\cite{Shewchuk:1994}. 
We enforced this stringent stopping criterion in order to approach the Moore-Penrose pseudoinverse solutions \cite{BarrettBook}. 
Note that the LS objectives, i.e, $ \Vert \mathbf u - \mathbf H_{\rm int} \boldsymbol\alpha_{\rm int} \Vert^2$ and $ \Vert \tilde{\mathbf u} - \tilde{\mathbf H}_{\rm KB} \boldsymbol\alpha_{\rm KB} \Vert^2$, were monotonically decreasing during the iteration. 
Even though  the images were reconstructed from noiseless data, one observes that artifacts are present (see Fig.~\ref{fig:noiselessZ}). 
These artifacts are due to the errors in the system matrices as well as the responses of the system matrices to the errors. 
These results suggest that $\tilde{\mathbf H}_{\rm KB}$ more accurately approximates the true underlying C-D imaging model, i.e., Eqn.~\eqref{eqn:CDmodel}, than does $\mathbf H_{\rm int}$. 

The residual of the cost functions decays faster in general when $\tilde{\mathbf H}_{\rm KB}$ is employed as shown in Fig.~\ref{fig:convergence}. 
It took $2675$ and $1782$ iterations to achieve the stopping criterion by use of $\mathbf H_{\rm int}$ and $\tilde{\mathbf H}_{\rm KB}$, respectively, suggesting a faster convergence rate by use of  $\tilde{\mathbf H}_{\rm KB}$ as predicted by the SVD analysis in Sect.~\ref{sect:svdresults}. 

As shown in Fig.~\ref{fig:mse_vs_niter}-(a), the minimal MSE appeared at the $37$-th and the $68$-th iteration by use of $\mathbf H_{\rm int}$  and $\tilde{\mathbf H}_{\rm KB}$ respectively, far before the final convergence. 
Images corresponding to the minimal MSEs are displayed in Fig.~\ref{fig:optmnoiseless}. 
The MSE of the image reconstructed by use of $\tilde{\mathbf H}_{\rm KB}$ (${\rm MSE}=0.90\times10^{-3}$) is about $90.0\%$ of that by use of $\mathbf H_{\rm int}$ (${\rm MSE}=1.00\times10^{-3}$). 
Even though the difference in MSE is insignificant, it can be observed that the image corresponding to $\mathbf H_{\rm int}$ (Fig.~\ref{fig:optmnoiseless}-(a)) contains more ripple artifacts than does the image corresponding to $\tilde{\mathbf H}_{\rm KB}$ (Fig.~\ref{fig:optmnoiseless}-(b)).
This observation is especially evident in the slowly-varying region as shown in Fig.~\ref{fig:optmnoiseless}-(c).
It is also interesting to note that $\tilde{\mathbf H}_{\rm KB}$ results in a larger overshoot in the region containing a small sharp structure (Fig.\ \ref{fig:optmnoiseless}-(d)), which is consistent with those observations made in previous studies of
 KB function-based image reconstruction \cite{Lewitt}. 
However, the circular shape of the small structure is better preserved by use $\tilde{\mathbf H}_{\rm KB}$ (see the reference in box-0 in Fig.~\ref{fig:noiselessZ}-(a)). 
In summary, $\tilde{\mathbf H}_{\rm KB}$ resulted in more accurate reconstruction than did $\mathbf H_{\rm int}$. 

It is notable that the minimal MSE defined in the plane of $z=0$ implies little on the accuracy of other regional MSE's as shown in Fig.~\ref{fig:mse_vs_niter}. 
As expected, all regional ${\rm MSE}$'s increase after initially declining because the errors in approximating the true C-D model (i.e.  Eqn.~\eqref{eqn:CDmodel}) with the system matrices are amplified during iterations and present as artifacts in the reconstructed images.  
However, the regional ${\rm MSE}$'s corresponding to $\mathbf H_{\rm int}$ increase more rapidly than do those corresponding to $\tilde{\mathbf H}_{\rm KB}$, suggesting $\tilde{\mathbf H}_{\rm KB}$ is numerically more stable. 
Also, the minimal values of various regional $\rm MSE$'s corresponding to $\tilde{\mathbf H}_{\rm KB}$ are in general smaller than those corresponding to $\mathbf H_{\rm int}$. 
This observation is especially evident in the uniform and slowly-varying ROIs (see Fig.~\ref{fig:mse_vs_niter}-(b) and -(c) respectively). 
These observations hold true for all $64$ realizations of $\underline A(\mathbf r)$.  
The $\rm EMSE$'s given in Table \ref{tab:emse} further confirm that images reconstructed by use of $\tilde{\mathbf H}_{\rm KB}$ are more accurate than those by use of $\mathbf H_{\rm int}$. 

\subsection{Images reconstructed from an ensemble of noisy data}

An ensemble of noisy images were reconstructed by solving Eqns.\ \eqref{eqn:costfuncint} and \eqref{eqn:costfuncbcc} with Tikhonov regularization. 
We swept the values of regularization parameters $\beta_{\rm int}$ and $\beta_{\rm KB}$ within the ranges $[20,400]$ and $[20,1000]$, respectively. 
We set $\Delta{\rm_s} = 0.14$ mm, $\Delta{\rm_s^b} = 0.2$ mm, $a=0.28$ mm, $\gamma=10.4$, $m=2$, and $\Delta{\rm_d}=0.0175$ mm.  
Accordingly, $\hat{\boldsymbol\alpha}_{\rm int}$ and $\hat{\boldsymbol\alpha}_{\rm KB}$ are of dimensions $64^3$ and $(45^3\times2)$ respectively. 

Figures~\ref{fig:noisy_example} and \ref{fig:noisy_profile} show an example in which the $\rm MSE$ and average variance of images reconstructed by use of $\tilde{\mathbf H}_{\rm KB}$ are $87.8\%$ and $60.7\%$ of those by use of $\mathbf H_{\rm int}$ respectively, where the $\rm MSE$ and average variance were calculated in the plane of $z=0$. 
The results suggest that the images reconstructed by use of $\tilde{\mathbf H}_{\rm KB}$ are not only less biased but also less varying than those reconstructed by use of $\mathbf H_{\rm int}$.
This observation holds true independently of the choice of regularization parameter as shown in Fig.~\ref{fig:mse_vs_var}-(a). 
Figure~\ref{fig:mse_vs_var}-(a) suggests that, for any choice of $\beta_{\rm int}$, there exists a $\beta_{\rm KB}$ such that images reconstructed by use of  $\tilde{\mathbf H}_{\rm KB}$ are more accurate as well as less varying among realizations.
Since they were calculated between the phantom and mean images, the $\rm MSE$s describe image bias averaged over ROIs. 
Within various ROIs, images reconstructed by use of $\tilde{\mathbf H}_{\rm KB}$ are always less biased than those reconstructed by use of $\mathbf H_{\rm int}$ when both are at the same variance level except for the region containing the small sharp structures (See Fig.~\ref{fig:mse_vs_var}). 
In addition, when $\beta_{\rm int}$ and $\beta_{\rm KB}$ took large values, the difference between the performace of $\tilde{\mathbf H}_{\rm KB}$ and $\mathbf H_{\rm int}$ is less obvious. 
These observations are also consistent with those observed in other imaging modalities \cite{Fessler:blob94,PETRef,wang20043d}.   

\subsection{Experimental Results}

The optimal performance of $\mathbf H_{\rm int}$ and $\tilde{\mathbf H}_{\rm KB}$ is displayed in Figs.~\ref{fig:exp_single_realization}, \ref{fig:exp_mean_var} and \ref{fig:exp_profile}, where the optimal performance is defined to be the case in which the $\rm MSE$ of the average reconstructed image is minimized by the optimal regularization parameter values. 
The optimal regularization parameters were estimated to be $\beta_{\rm int}=0.03$ and $\beta_{\rm KB}=0.1$ by a brute-force search.
Note that the $\rm MSE$ was defined in the plane of $y=1.4$ mm using a display grid of spacing $\Delta{\rm_d}=0.0175$ mm. 

The $\rm MSE$ of the mean image, averaged over the $64$ measurements, reconstructed by use of $\tilde{\mathbf H}_{\rm KB}$ (${\rm MSE} = 1.43\times10^{-4}$) is about $53\%$ of that by use of $\mathbf H_{\rm int}$ ($\rm MSE = 2.68\times10^{-4}$). 
Pixelated edges are observed in both the mean image (see Fig.~\ref{fig:exp_mean_var}-(a)) and the image reconstructed from a single measurement (see Fig.~\ref{fig:exp_single_realization}-(b)) by use of $\mathbf H_{\rm int}$.  
In contrast, the pixelation effect is much less noticeable in the images reconstructed by use of $\tilde{\mathbf H}_{\rm KB}$ (see Figs.~\ref{fig:exp_single_realization}-(c) and \ref{fig:exp_mean_var}-(b)).
This is expected since the choice of $\{\psi_n^{\rm KB}(\mathbf r)\}$ constrains $\hat A_{\rm KB}(\mathbf r)$ to be differentiable in space. 
Further, profiles of the reconstructed images (see Fig.~\ref{fig:exp_profile}) indicate a notable quantitative error in the images reconstructed by use of $\mathbf H_{\rm int}$. 
This observation is consistent with our computer-simulation results that suggest that slowly varying regions can be more accurately reconstructed by use of $\tilde{\mathbf H}_{\rm KB}$ (see Fig.~\ref{fig:mse_vs_niter}-(c) and Table \ref{tab:emse}). 
In addition, one observes spatially dependent variances among images reconstructed from $64$ measurements as shown in Fig.~\ref{fig:exp_mean_var}-(c) and -(d). 
Specifically, the variance maps contain structural patterns, suggesting object dependent noise statistics \cite{Mandelis:PATSNR}. 
At the optimal performance, the average variance corresponding to $\tilde{\mathbf H}_{\rm KB}$ ($\sim6.14\times 10^{-4}$) is about $78\%$ of that corresponding to $\mathbf H_{\rm int}$ ($\sim7.83\times 10^{-4}$). 
This observation is predicted by the singular value analysis in Sect.~\ref{sect:svdresults}. 

We estimated the optical energy densities within two ROIs marked in Fig.~\ref{fig:exp_single_realization}-(a), where the true energy densities estimated from the reference image were $0.64$ and $0.45$ in arbitrary units, respectively, for ROI-A and ROI-B. 
Both ROIs are of dimension $1\times1$ mm$^2$.
We swept the values of $\beta_{\rm int}$ and $\beta_{\rm KB}$ within the ranges $[0,0.15]$ and $[0,1.0]$ respectively. 
Within these ranges, the plots corresponding to $\tilde{\mathbf H}_{\rm KB}$ are always below the plots corresponding to $\mathbf H_{\rm int}$ as shown in Fig.~\ref{fig:exp_biasvsvar}.
The results suggest that optical energy densities can be more accurately and stably estimated by use of of $\tilde{\mathbf H}_{\rm KB}$ than by use of $\mathbf H_{\rm int}$.

\section{Discussion}


The KB function-based imaging model investigated in this work
 generalizes the uniform-spherical-voxel-based imaging model we proposed earlier \cite{TMI:transmodel,Kun:PMB}.
This generalization maintains the convenience in modeling the finite aperture size effect of ultrasonic transducers (see Eqn.~\eqref{eqn:blobsysmatrix}) while reducing computation by a factor of $\sqrt{2}$ with the use of an equivalent BCC grid.
Computer-simulation and experimental results have demonstrated that the KB function-based imaging model is, in general, not only quantitatively more accurate but also numerically more stable than a conventional linear-interpolation-based imaging model. 
By use of iterative image reconstruction algorithms based on KB function, absorbed optical energy densities can be more accurately estimated with smaller variances.

The KB function-based imaging model possesses at least two limitations.
First, if the object contains fine sharp structures possessing a dimension that
 is smaller than the KB function radius, the KB function-based imaging model may lead
 to a overshoot in the reconstructed images as shown in Fig.~\ref{fig:optmnoiseless}-(d). 
Second, the computational complexity for KB  function-based iterative image reconstruction is, in general, 
 higher than that for interpolation-based iterative image reconstruction.  
As described below, for the application presented in this study, the computational time required to
 complete one iteration was approximately
 $50\%$  longer for the KB function-based imaging model than for the linear-interpolation-based imaging model. 

Even if ultrasonic transducers can be accurately approximated as point-like, the KB function-based imaging model still outperforms a conventional linear interpolation model in certain aspects.
For many object functions of practical interest, the KB function-based imaging model can more accurately approximate the true C-D model, i.e, Eqn.~\eqref{eqn:CDmodel}, than does the linear-interpolation-based imaging model.
Particularly in regions containing smooth structures, use of the KB function-based imaging model can significantly improve the accuracy of reconstructed images (see Fig.~\ref{fig:mse_vs_niter}-(b) and -(c) and Table \ref{tab:emse}). 
Moreover, the KB function-based imaging model appears to be more robust to random noise as predicted by the singular value spectra (see Fig.~\ref{fig:svd}).  
These advantages are due to the fact that the KB function-based representation constrains reconstructed images to be spatially differentiable as well as the fact that the KB function-based system matrix is analytically calculated with no numerical approximations on the time derivative term \cite{Qiaofeng:12,Ntziachristos:QPAT2012}. 
Therefore, we believe that the superior performance of the KB function-based imaging model will persist even if different optimization algorithms or different linear-interpolation-based imaging models \cite{GPaltauf:2002,AnastasioTATHTfeasible,Jin:2009,German3D:12} are employed.

To our knowledge, this is the first study in which iterative image reconstruction algorithms were evaluated by use of a parameter estimation task in OAT \cite{Patrick13:hotelling}. 
Task-based imaging quality assessment is seldom employed in OAT studies \cite{Patrick13:hotelling}.
An important reason is that the necessary statistical studies \cite{BarrettBook} are in general computationally burdensome, particularly if iterative image reconstruction algorithms are of interest. 
Our GPU-based implementations \cite{Kun:GPU} greatly accelerate the computation, increasing the feasibility of task-based image quality assessment. 
On the platform consisting of dual quad-core CPUs with a clock speed $3.30$ GHz, each iteration, running on a single NVIDIA Tesla K20 GPU,  took $33$ and $45$ seconds to process the experimental data by use of the linear-interpolation- and KB 
 function-based imaging models, respectively. 
The number of iterations required varied among $20$ to $200$, depending on regularization parameter values. 
Our task-based image quality assessment study is far from comprehensive, 
but it is interesting to observe the dependence in the noise pattern on the image reconstruction algorithms (see Fig.~\ref{fig:exp_mean_var}-(c) and -(d)). 
How the noise pattern affects tasks such as tumor detection remains an interesting and open topic for future studies \cite{BarrettBook,Patrick13:hotelling}.


%

\appendices
\section{Derivation of the pressure generated by radially symmetric expansion functions}
In a homogeneous medium in three-dimensions, the pressure, $\tilde{p}({\bf{r}},f)$ induced via the photoacoustic effect is given by 
\begin{eqnarray}
\tilde{p}({\bf{r}},f) = \frac{\hat\jmath f\beta}{2 C_p} \int \mathrm{d} \mathbf r'\, \frac{\exp\left(-\hat\jmath k|{\bf{r}}-{\bf{r'}}| \right)}{|{\bf{r}}-{\bf{r'}}|}A({\bf{r'}})
\end{eqnarray}
where $f$ is the frequency and $k=2\pi f/c_0$. Suppose the source is described by
a spherically symmetric function, namely, $A({\bf{r}}) = a(r)$, where $r \in \mathbb R^+$.  The pressure is then given by 
\begin{eqnarray}
\tilde{p}({\bf{r}},f) &=&  \frac{\hat\jmath f \beta}{2 C_p} \int \mathrm{d} \mathbf r'\, \frac{a(r)}{|{\bf{r}}-{\bf{r'}}|} \exp(-\hat\jmath k |{\bf{r}}-{\bf{r'}}|) \\
&=& \label{temp:eq}  \frac{- \beta c_0}{2 C_p} \frac{\exp(-\hat\jmath k r)}{r}\\ \nonumber &\times& \int_0^\infty \mathrm{d}r'\,r'a(r') [\exp(\hat\jmath k r')-\exp(-\hat\jmath k r')].
\end{eqnarray}
The last step was performed by evaluating the integral in spherical coordinates over the azimuthal and polar coordinates.
Introducing the auxiliary function
\begin{eqnarray}
\bar{a}(r) = \left\{ \begin{array}{l l} a(r) & r\geq 0 \\ a(-r) & r<0, \end{array} \right.
 \end{eqnarray}
the expression in Eq.~(\ref{temp:eq}) can be simplified to
\begin{eqnarray} \nonumber
\tilde{p}({\bf{r}},f) &=& -\frac{\beta c_0}{2C_P}\frac{\exp(-\hat\jmath kr)}{r}\int_{-\infty}^\infty \mathrm{d}r'\,r'\bar{a}(r') \exp(\hat\jmath kr') \\ \nonumber
&=&  -\frac{\beta c_0}{2C_P}\frac{\exp(-\hat\jmath kr)}{r} \frac{c}{-\hat\jmath}\frac{\partial}{\partial \omega}\int_{-\infty}^\infty \mathrm{d}r'\,\bar{a}(r') \exp(\hat\jmath k r') \\
&=&\label{F_fin}  \frac{-\hat\jmath\beta c^2_0}{4\pi C_p}\frac{\exp(-\hat\jmath k r)}{r} \frac{\partial}{\partial f} \mathcal{A}(2\pi f/c_0).
\end{eqnarray}
where $\mathcal{A}(k)$ is the one-dimensional Fourier transform of $\bar{a}(r)$ and the derivative identity for Fourier transforms was used.  {Equation~(\ref{F_fin}) can be used to calculate the pressure induced by any integrable
and radially symmetric expansion function. } In the specific case that $a(r)$ represents a KB function, the Fourier transform of the KB  function of order $m$ can be found in $p$ spatial dimensions via Sonine's second integral formula \cite[see Sec. 12.13]{watson1995treatise} as described in Lewiit \cite{lewitt1990multidimensional}:
\begin{eqnarray}
\mathcal{A}_{m}^{(p)}(k) = \frac{(2\pi)^{p/2} a^p \gamma^m}{I_m(\gamma)}\frac{J_{p/2+m}(\sqrt{k^2a^2-\gamma^2})}{(\sqrt{k^2a^2-\gamma^2})^{p/2+m}},
\end{eqnarray}
where $\mathcal{A}_{m}^{(p)}(k)$ is the spatial Fourier transform of a KB function of order $m$ in $p$-dimensions.
Substituting the form for the Fourier transform of the KB function into Eq.~(\ref{F_fin}) for $p =1$ dimensions, the pressure generated by a KB function centered at the origin is given by the temporal frequency domain expression: 
\begin{equation}\label{eqn:pfblob}
  \tilde{p}({\bf{r}},f) = - \frac{\hat\jmath 2\pi  f a^3 \beta}{C_p I_m(\gamma)}\frac{\exp(-\hat\jmath kr)}{r} \frac{j_{m+1}(\sqrt{k^2a^2-\gamma^2})}{(k^2a^2-\gamma^2)^{(m+1)/2}},
\end{equation}
where, again,  $k = 2\pi f/c_0$ and 
\begin{eqnarray}
j_m(x) = \sqrt{\frac{\pi}{2x}}J_{m+1/2}(x)
\end{eqnarray}
is the spherical Bessel function of order $m$.

Note that taking the inverse Fourier transform of the expression for the pressure in  Eq.~(\ref{F_fin}) gives an exact expression for the time-domain pressure generated by a spherically symmetric source:
\begin{eqnarray}
p({\bf{r}},t) &=& \frac{-\hat\jmath\beta c^2_0}{2 C_p}\frac{1}{r} \int \mathrm{d} f\,\exp(\hat\jmath 2\pi f t) \exp(-\hat\jmath kr) \frac{\partial}{\partial f} \mathcal{A}(2\pi f/c_0) \\
&=& \frac{\beta c_0^2}{2C_p}\frac{r-c_0t}{r}\bar{a}(r-c_0t) 
\end{eqnarray}
which agrees with previous results \cite{diebold2009photoacoustic}.

\section*{Acknowledgment}
This research was supported in part by NIH awards  EB010049, CA167446, EB016963, and EB014617.

\ifCLASSOPTIONcaptionsoff
  \newpage
\fi

\bibliographystyle{IEEEtran}

\newpage
\section*{Tables}

\begin{table}[h]
\caption{\label{tab:phantom} Parameters of the random numerical phantom}
\centering
\begin{tabular}[h]{c |ll |ll |ll |l}
\hline\hline
         &\multicolumn{2}{c|}{$(\underline{x}_i,\underline{y}_i,\underline{z}_i)$}
         &\multicolumn{2}{c|}{$\underline{R}_i$}
         &\multicolumn{2}{c|}{$\underline{A}_i$}
         &{$G_i(\mathbf r)$}  \\
index    &mean [mm] & STD [mm] & mean [mm] & STD [\%]&mean [a.u] & STD [\%] & FWHM [mm]\\[0.5ex]
\hline
$0$      &$(-0.57,-0.57,0)$ & $(0,0,0)$          & $3.50$  & $0$   & $0.30$  & $0 $ & $0$\\
$1$      &$(-0.57,-0.57,0)$ & $(0,0,0)$          & $3.00$  & $5.0$ & $-0.10$ & $20 $& $0.462$\\
$2$      &$(-2.10,-1.60,0)$ & $(0.30,0.30,0.30)$ & $0.50$  & $10$  & $0.50$  & $20 $& $0.154$\\
$3$      &$(-2.10, 0.46,0)$ & $(0.30,0.30,0.30)$ & $0.50$  & $10$  & $0.50$  & $20 $& $0.154$\\
$4$      &$(    0,-2.10,0)$ & $(0.30,0.30,0.30)$ & $1.00$  & $10$  & $0.30$  & $20 $& $0.154$\\
$5$      &$(-0.40, 0.06,0)$ & $(0.30,0.30,0.30)$ & $1.00$  & $10$  & $0.30$  & $20 $& $0.154$\\
$6$      &$( 0.40, 1.20,0)$ & $(0.30,0.30,0.30)$ & $0.16$  & $10$  & $0.80$  & $20$ & $0$\\
$7$      &$( 1.40,-0.40,0)$ & $(0.30,0.30,0.30)$ & $0.16$  & $10$  & $0.80$  & $20$ & $0$\\
$8$      &$( 1.20, 0.40,0)$ & $(0.30,0.30,0.30)$ & $0.16$  & $10$  & $0.80$  & $20$ & $0$\\
\hline\hline
\end{tabular}
\end{table}
\clearpage

\begin{table}[h]
\caption{\label{tab:emse} Ensemble mean-square errors within various ROIs (${\rm mean}\pm{\rm STD}$) in units of $\times 10^{-4}$}
\centering
\begin{tabular}[h]{c| c |c |c |c |c|c}
\hline\hline
         {System matrix}
         &{Plane $z=0$}
         &{Sharp small}
         &{Sharp large}
         &{Moderately blurred}
         &{Slowly varying}
         &{Uniform}  \\
\hline
$\mathbf H_{\rm int}$
& $7.69\pm1.26$
& $187\pm77.3$
& $23.8\pm0.222$
& $2.11\pm1.29$
& $0.420\pm0.0670$
& $1.66\pm0.798$\\
$\tilde{\mathbf H}_{\rm KB}$
& $6.80\pm1.11$
& $166\pm69.2$
& $23.2\pm0.108$
& $0.803\pm0.582$
& $0.284\pm0.0476$
& $0.411\pm0.186$\\
\hline\hline
\end{tabular}
\end{table}
\clearpage

\section*{Figures}

\begin{figure}[h]
\centering 
  \includegraphics[width=4.5in]{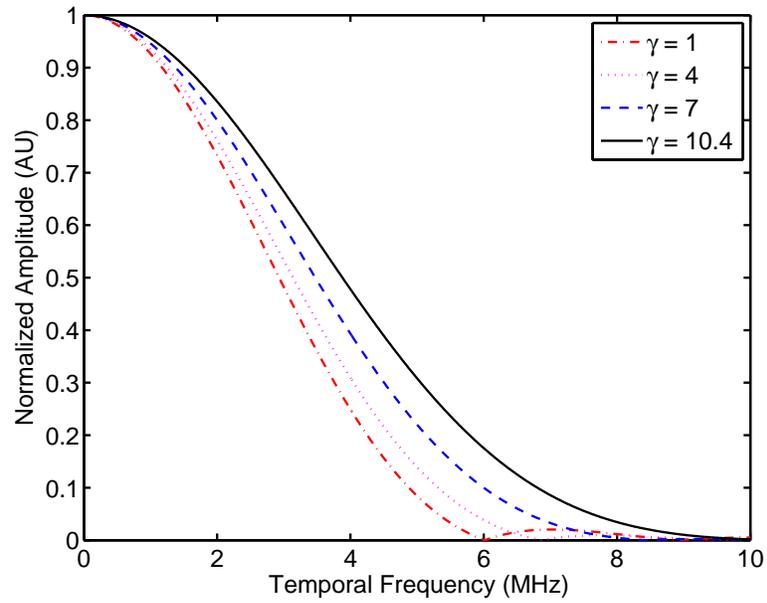}
  \caption{\label{blob_profiles}
Plots of the frequency content of the pressure resulting from a single KB function when $\gamma = 1$ (red dash-dots), $\gamma = 4$ (purple dots), $\gamma = 7$ (blue dashes) and $\gamma = 10.4$ (black solid line).  In this simulation, $m=2$ and $a = 0.28$ mm.
}
\end{figure}
\clearpage

\begin{figure}[h]
\centering
  \subfloat[]{{\includegraphics[width=1.5in]{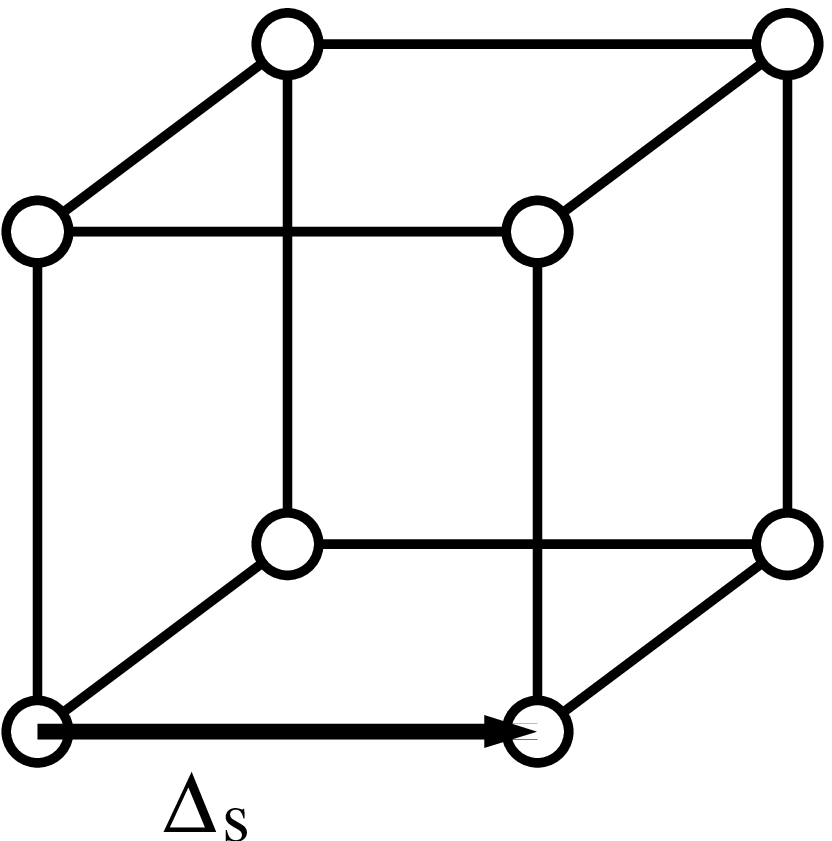}}}\hskip 0.5cm
  \subfloat[]{{\includegraphics[width=1.5in]{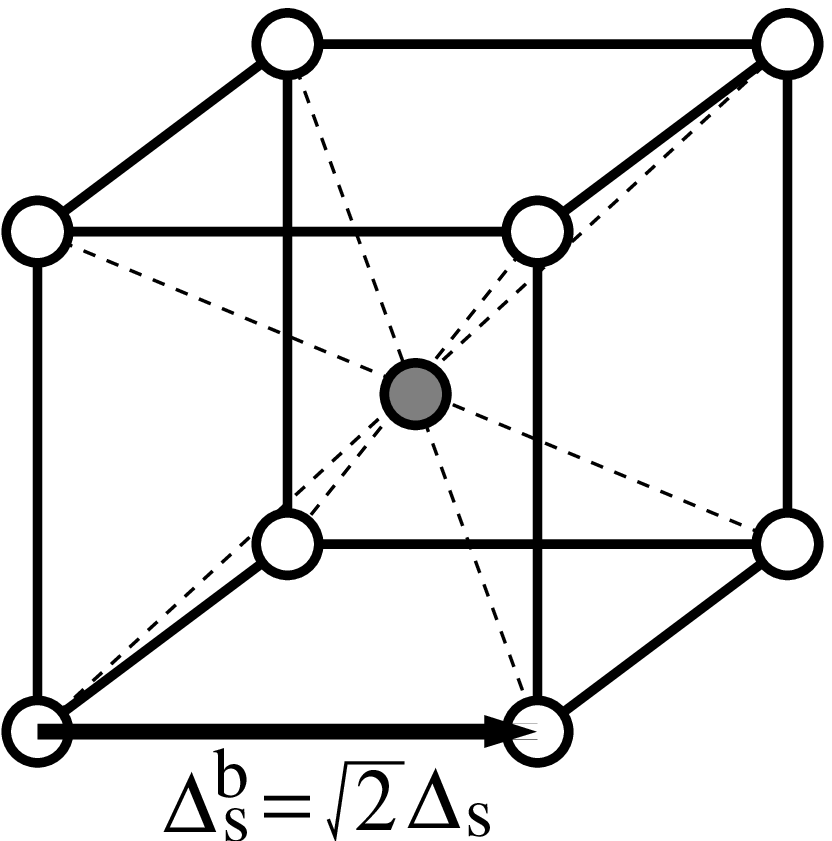}}}\hskip 0.5cm
  \subfloat[]{{\includegraphics[width=1.5in]{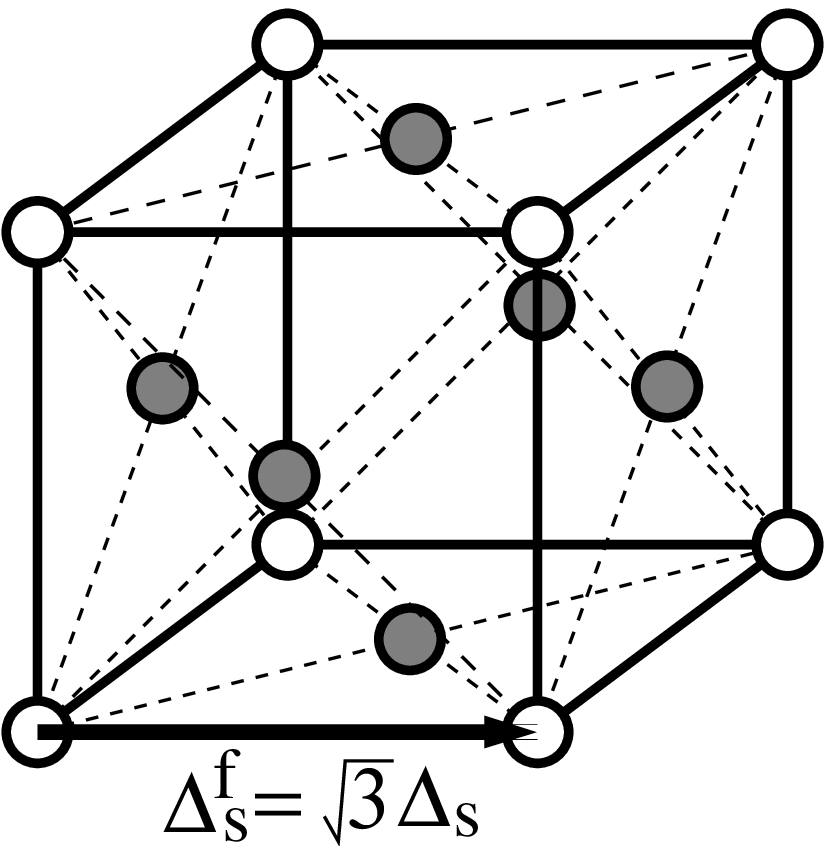}}}
\caption{\label{fig:gridstructure}
Sketches of three ``equivalent'' grid structures: (a) SC grid with spacing $\Delta{\rm_s}$, (b) BCC grid with spacing $\Delta{\rm_s^b}=\sqrt{2}\Delta{\rm_s}$ and (c) FCC grid with spacing $\Delta{\rm_s^f}=\sqrt{3}\Delta{\rm_s}$. }
\end{figure}
\clearpage

\begin{figure}[h]
\centering
\subfloat[]{\includegraphics[width=6cm]{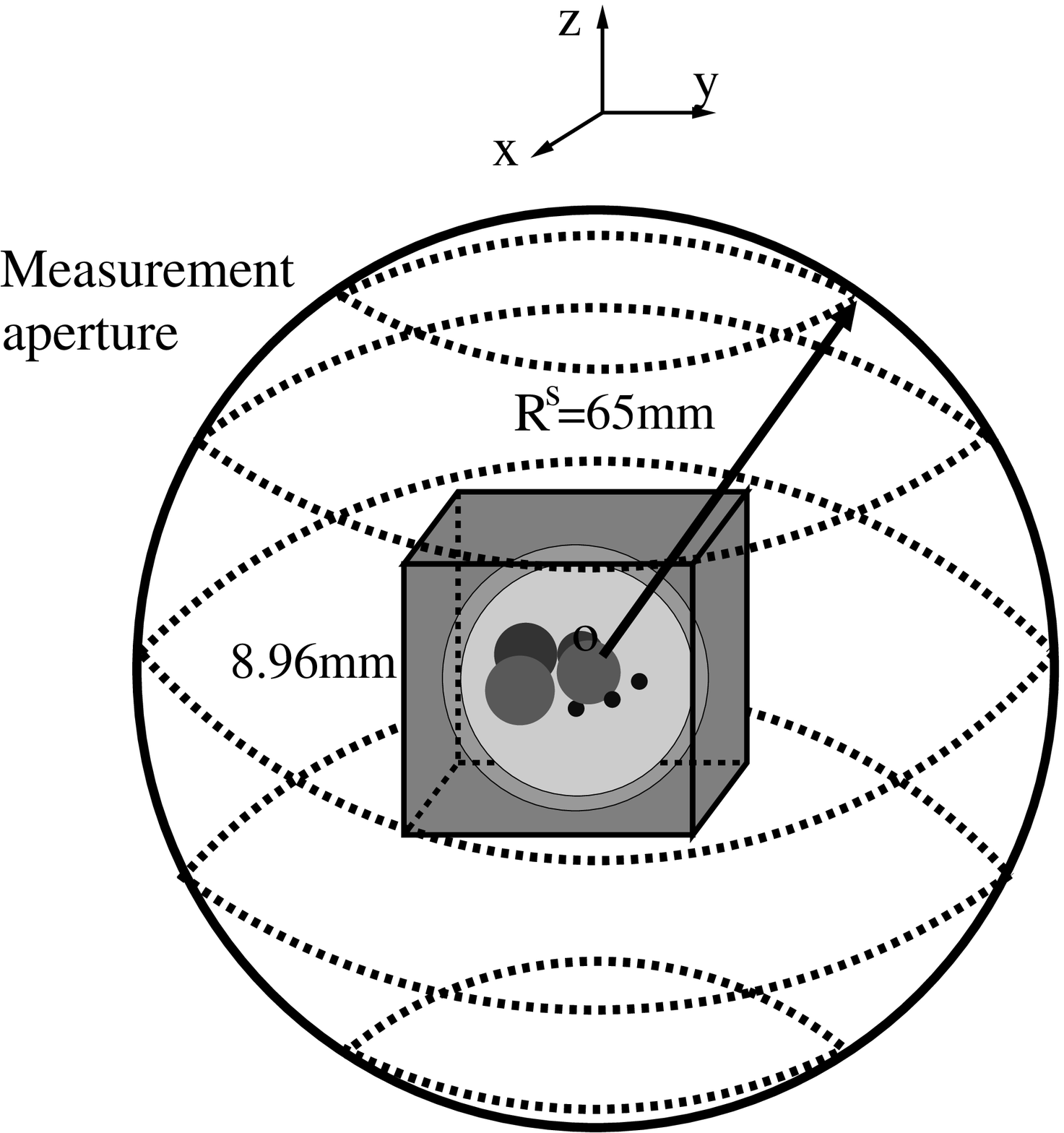}}
\hskip 1cm
\subfloat[]{\includegraphics[width=6.3cm]{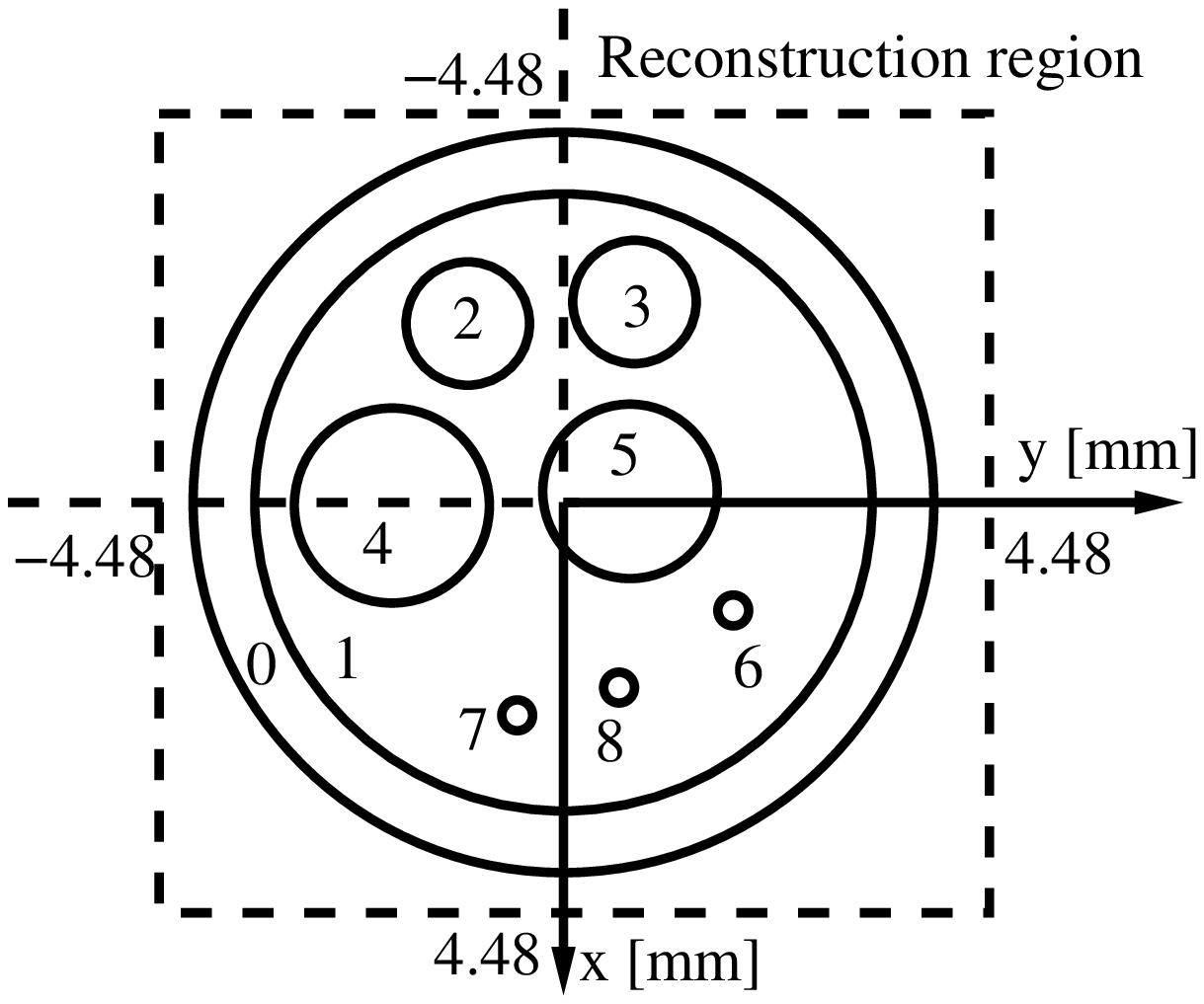}}
\caption{\label{fig:geo}(a) A schematic of the measurement geometry.
(b) A schematic of the 3D random object function in the plane $z=0$.}
\end{figure}
\clearpage

\begin{figure}[h]
\centering 
  \includegraphics[width=4.5in]{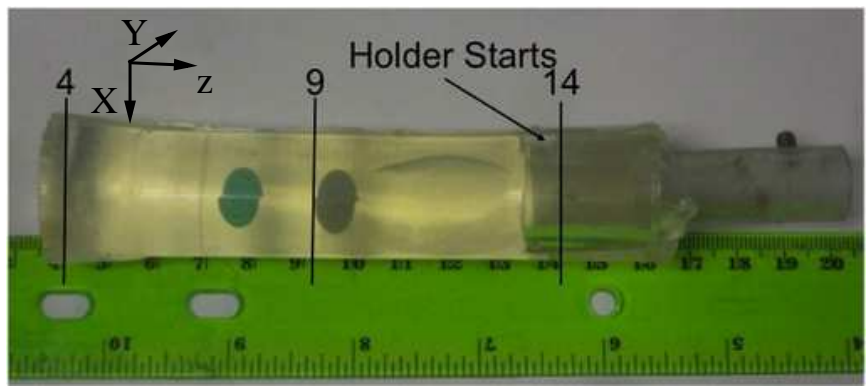}
  \caption{\label{fig:target}
Photograph of the experimental phantom. 
}
\end{figure}
\clearpage

\begin{figure}[h]
\centering 
  \includegraphics[width=5in]{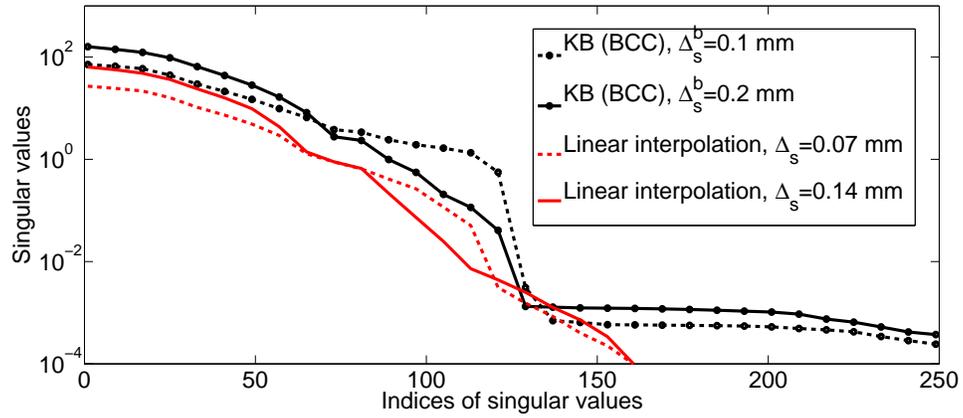}
  \caption{\label{fig:svd}
  Singular value spectra of the reduced-dimensional system matrices $\mathbf H_{\rm int}$ and $\tilde{\mathbf H}_{\rm KB}$ with grid spacing $\Delta{\rm_s}=0.07$ mm (or $\Delta{\rm_s^b}=0.1$ mm ) and $\Delta{\rm_s}=0.14$ mm (or $\Delta{\rm_s^b}=0.2$ mm).
}
\end{figure}
\clearpage

\begin{figure}[h]
\centering
  \subfloat[]{{\includegraphics[height=2.0in]{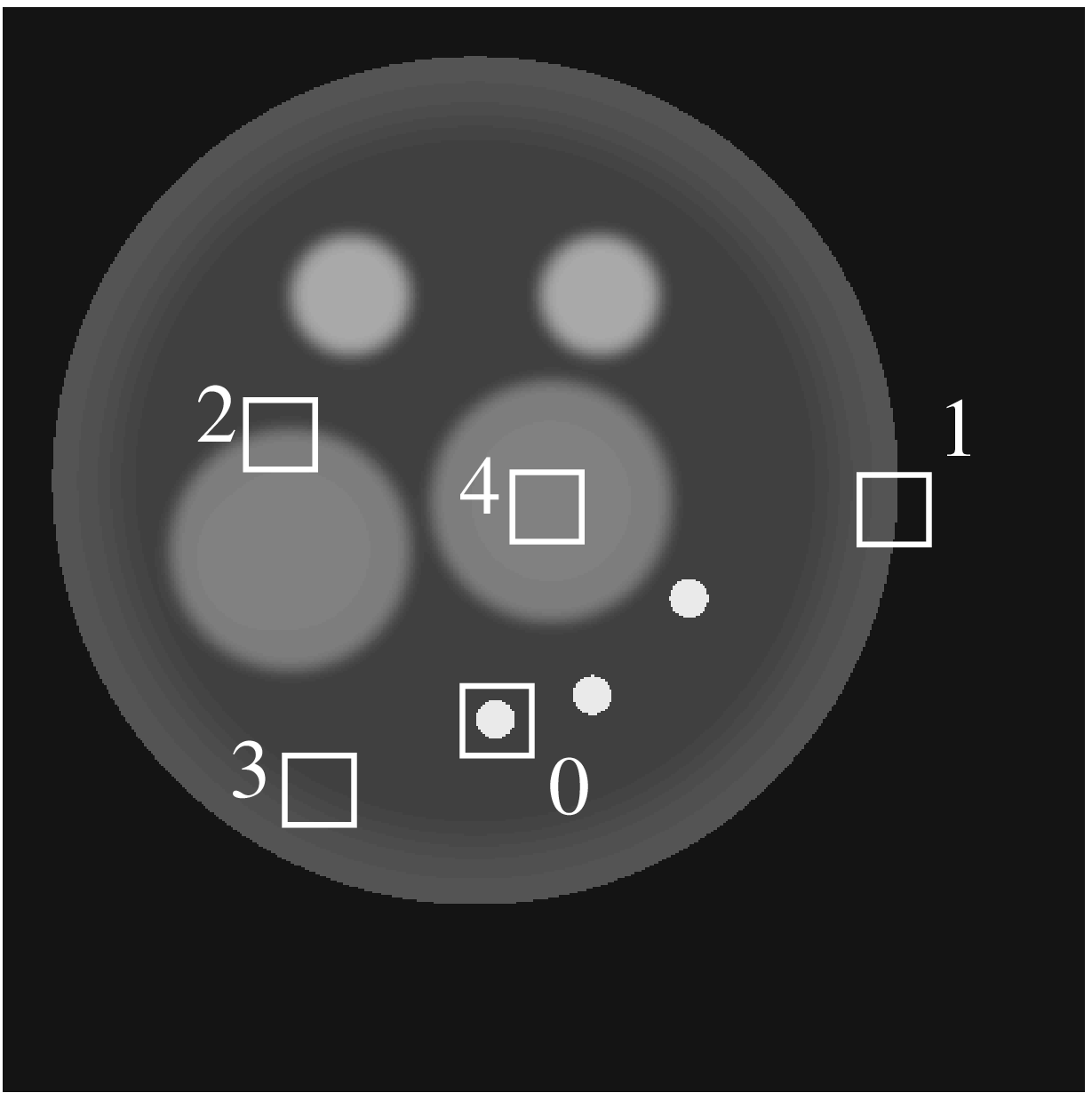}}}\hskip 0.1 cm
  \subfloat[]{{\includegraphics[height=2.0in]{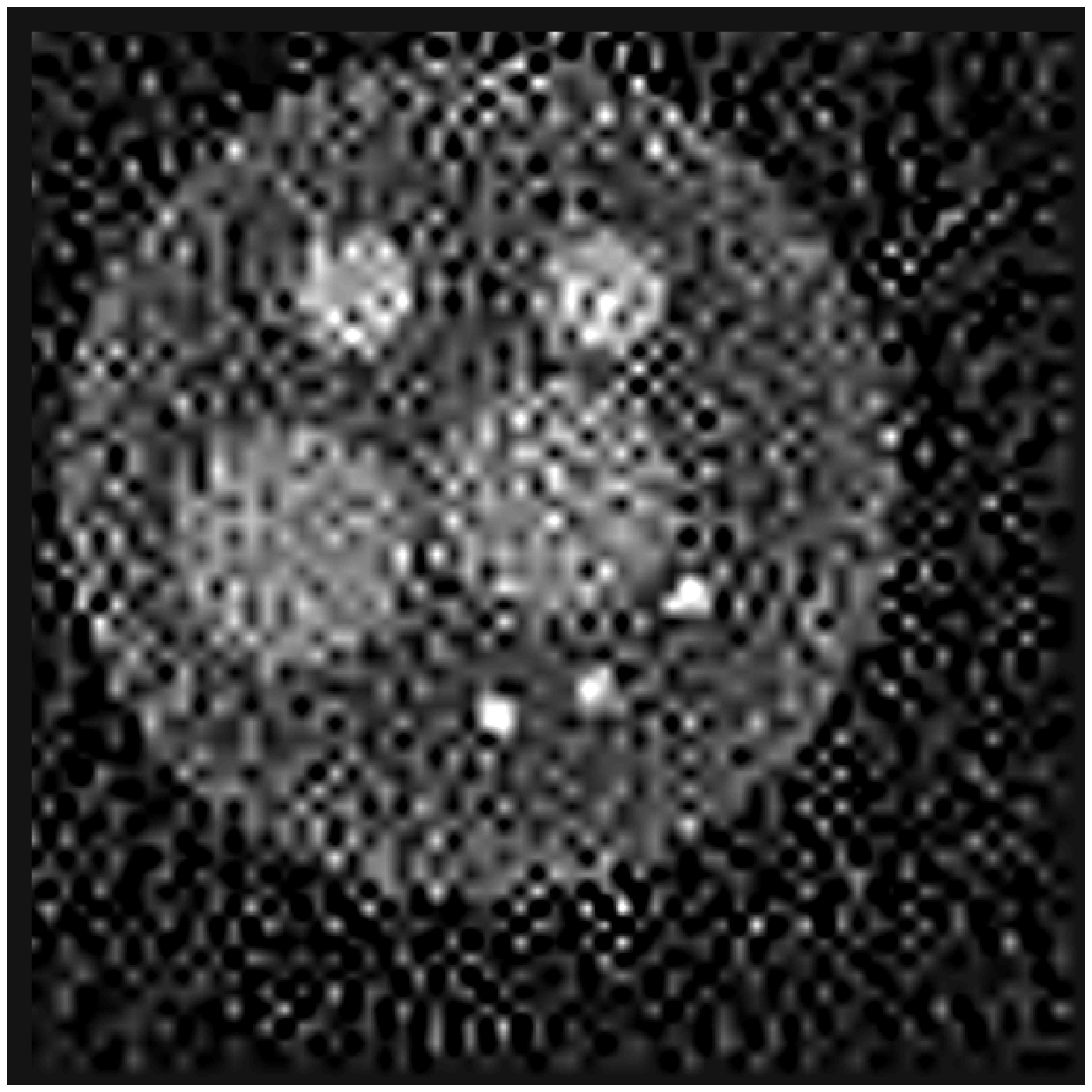}}}\hskip 0.1cm
  \subfloat[]{{\includegraphics[height=2.0in]{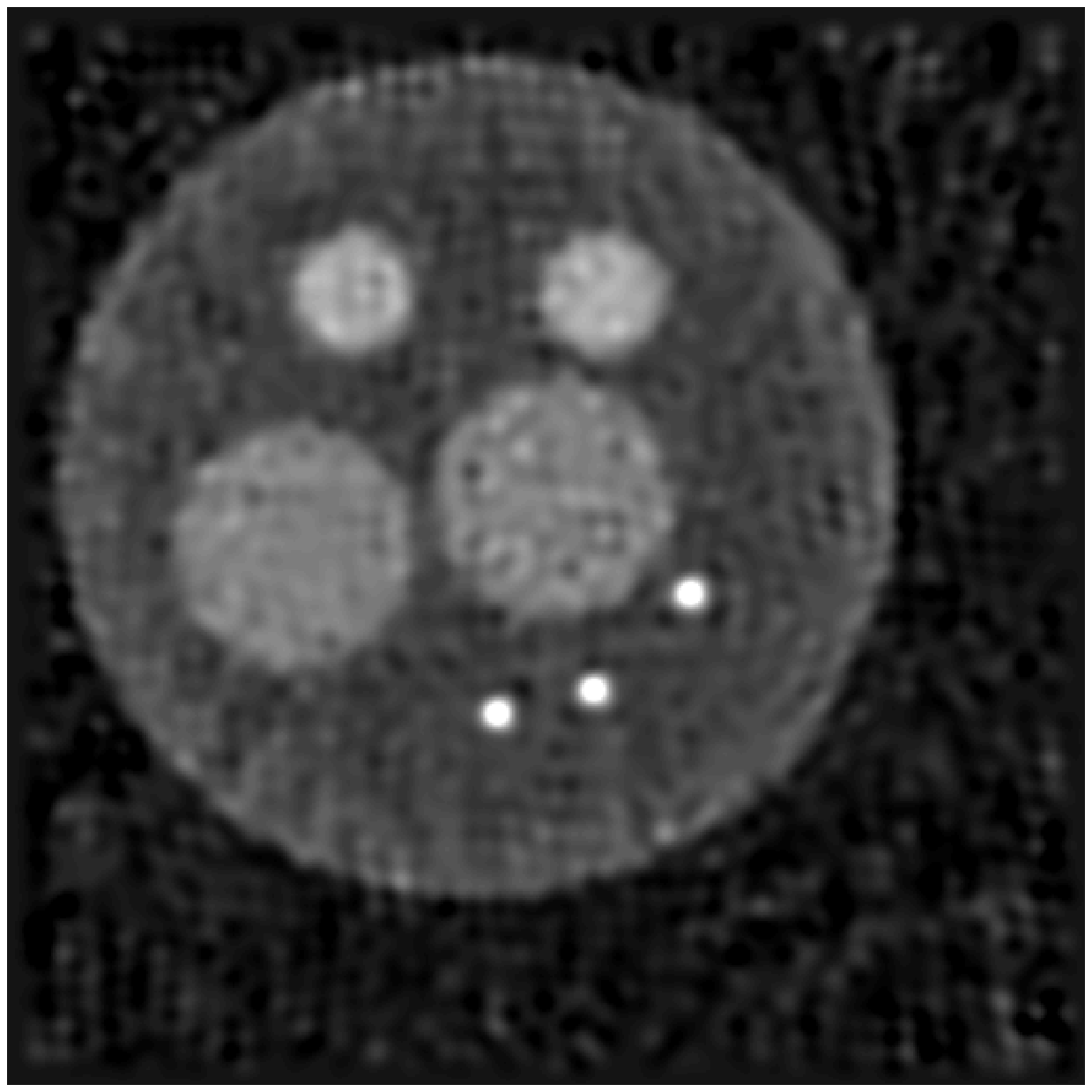}}}\hskip -0.1cm
  \includegraphics[height=2 in]{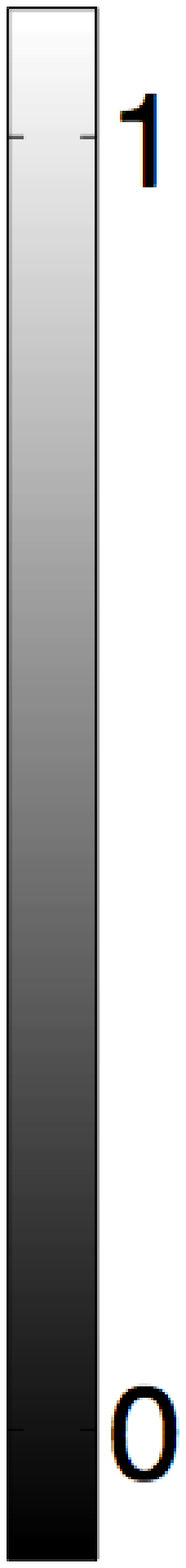}
\caption{\label{fig:noiselessZ}
Slices corresponding to the plane of $z=0$ through 
(a) the $3$D phantom $A^{(0)}(\mathbf r)$ 
and the $3$D images reconstructed by use of system matrices
(b) $\mathbf H_{\rm int}$ 
and 
(c) $\tilde{\mathbf H}_{\rm KB}$ from noise-free data.
The grayscale window is $[-0.1,1.1]$.  
Each reconstruction was terminated when the residual of the cost function was reduced to $0.01\%$ of its Euclidean norm. 
In panel (a),  the five ROIs used to calculate regional mean-square errors are contained inside white boxes.
}
\end{figure}
\clearpage

\begin{figure}[h]
\centering
  \includegraphics[width=4.0in]{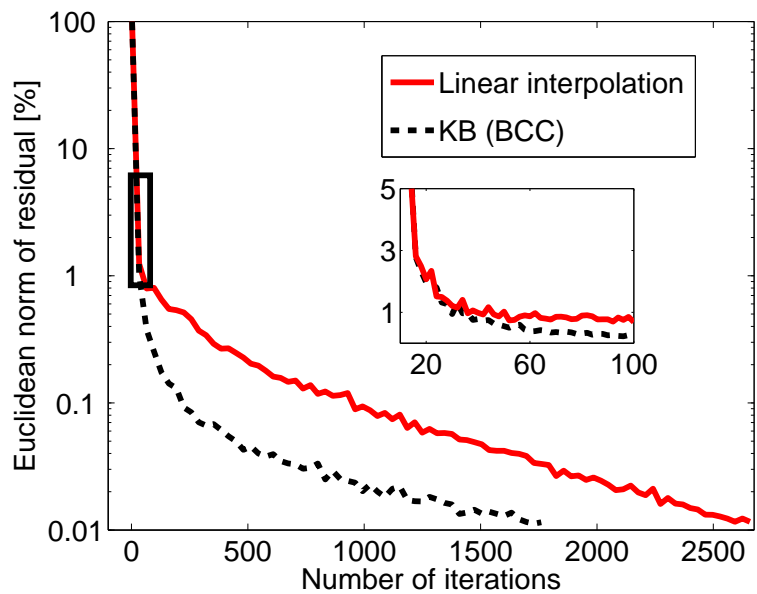}
\caption{\label{fig:convergence}
Decay of the residual of the cost function against the number of iterations. 
The subplot shows the decay in linear coordinates corresponding to the portion contained in the black box.
}
\end{figure}
\clearpage

\begin{figure}[h]
\centering
  \subfloat[]{{\includegraphics[width=2.5in]{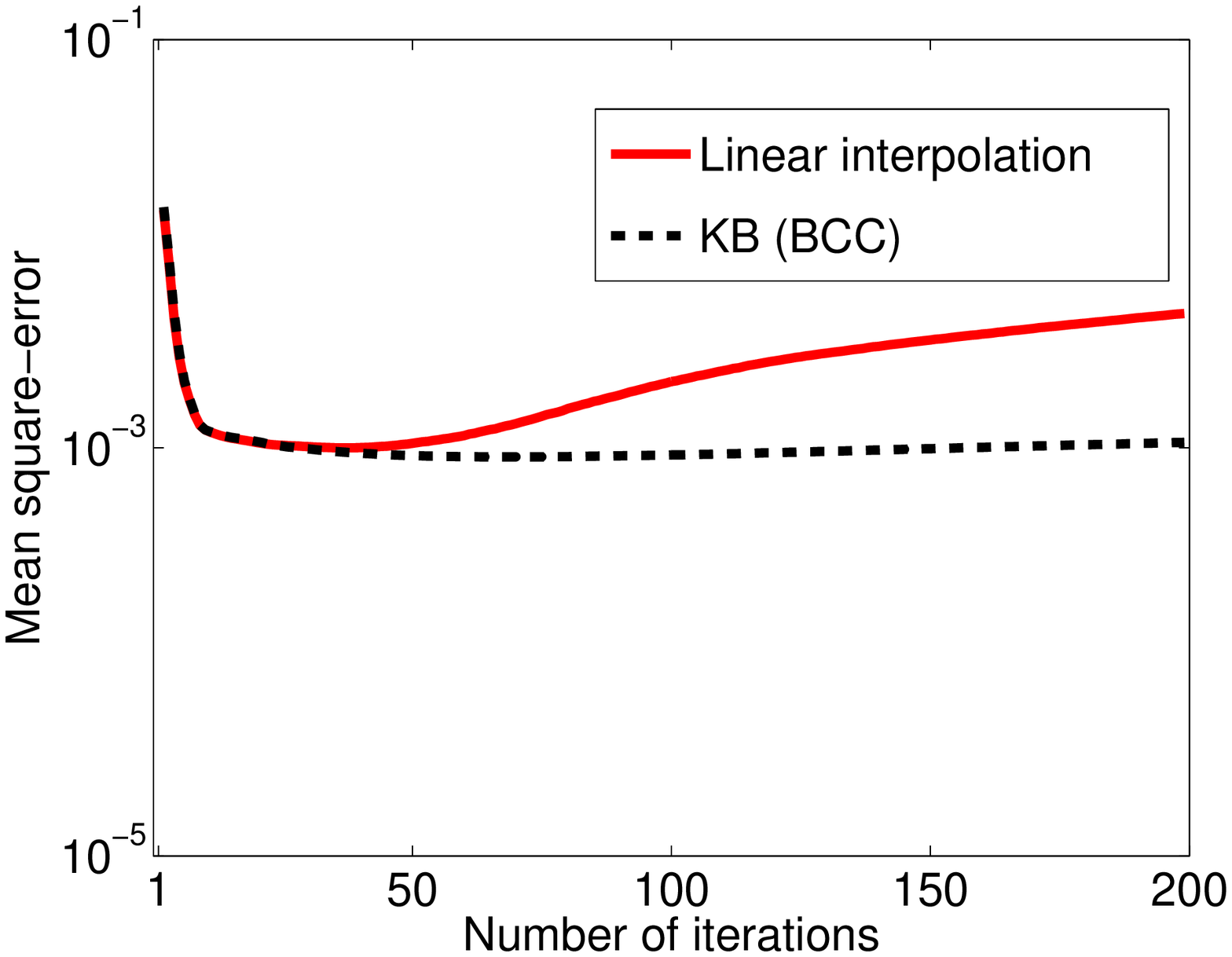}}}\hskip 0.5cm
  \subfloat[]{{\includegraphics[width=2.6in]{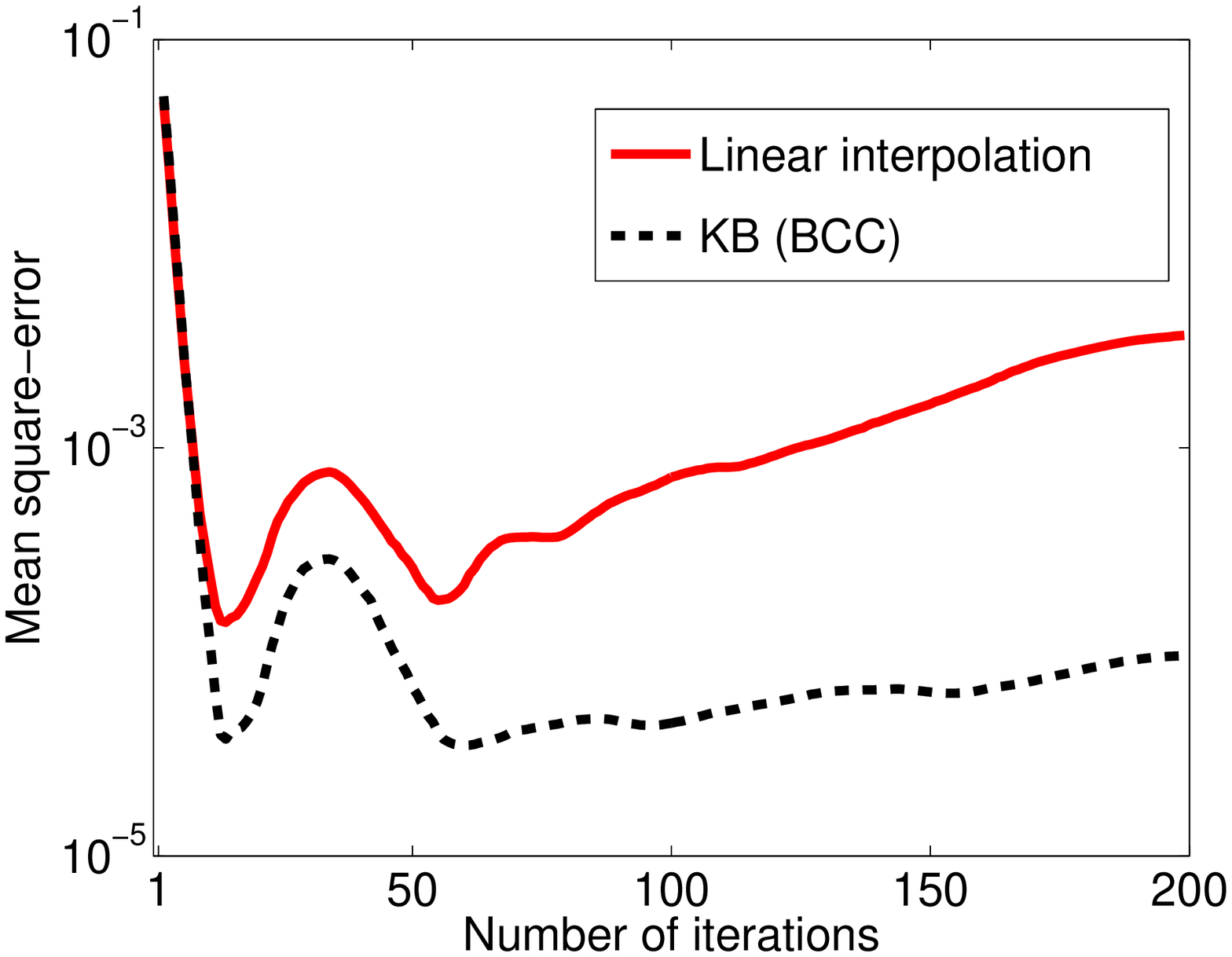}}}\\
  \subfloat[]{{\includegraphics[width=2.5in]{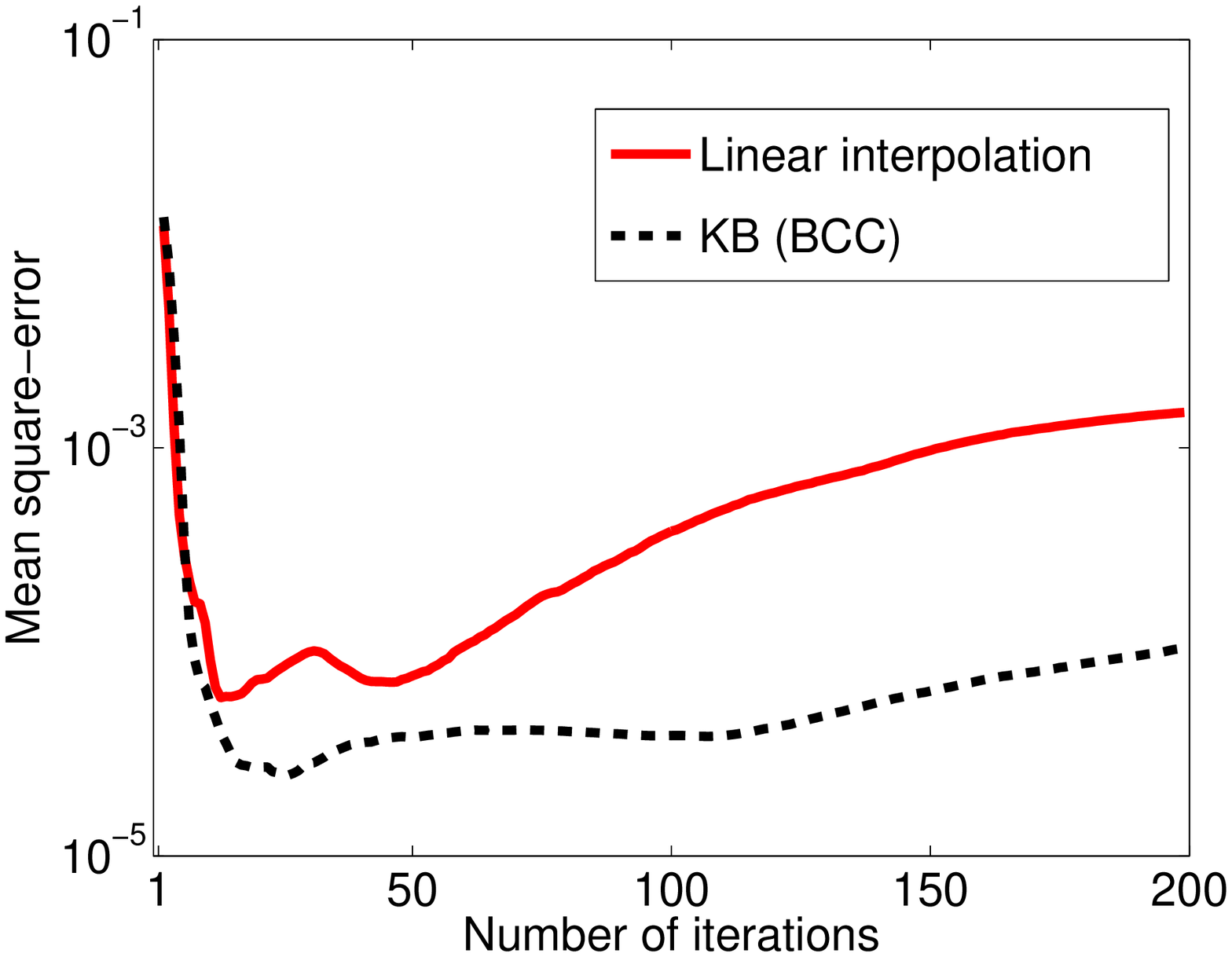}}} \hskip .5cm
  \subfloat[]{{\includegraphics[width=2.5in]{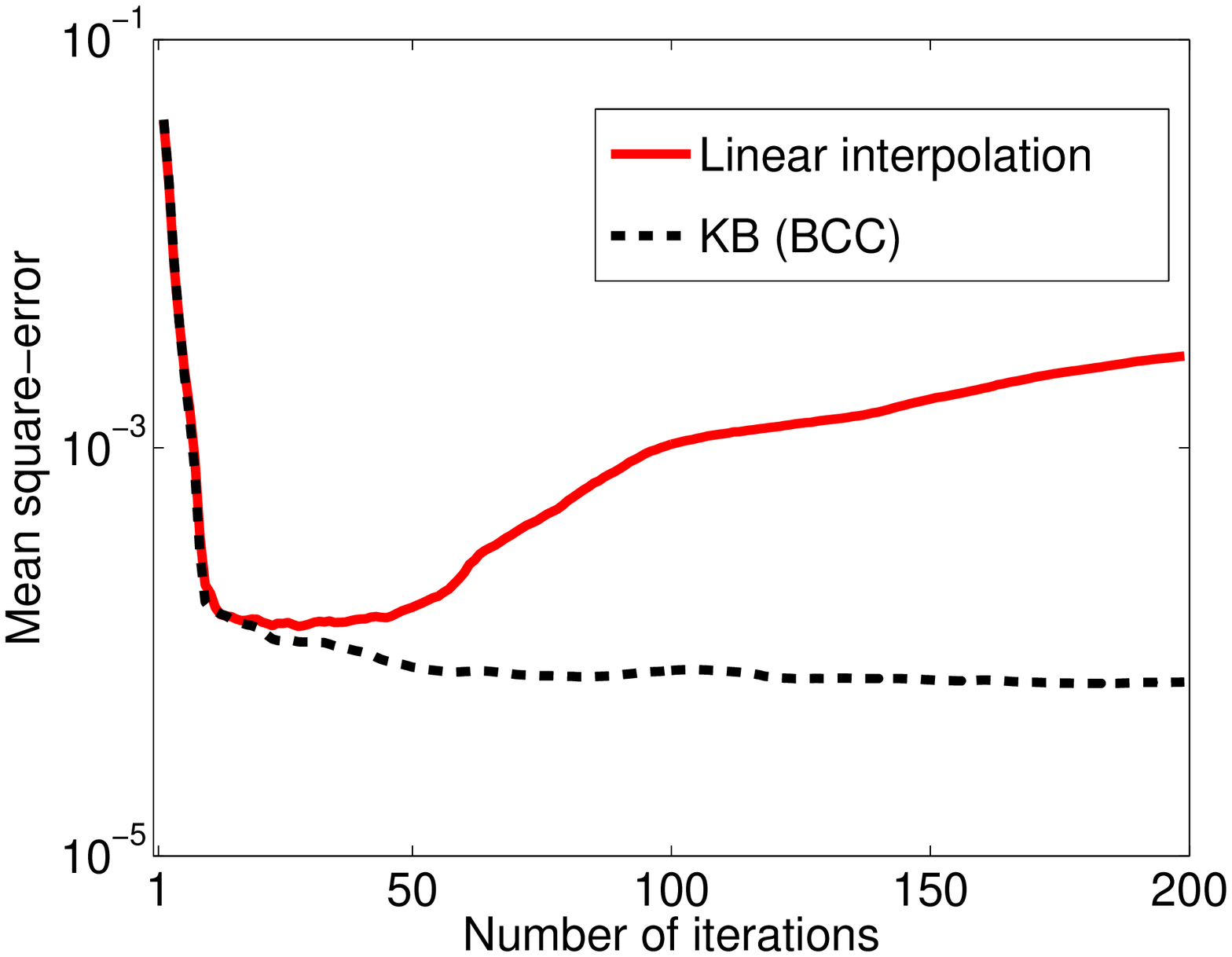}}}\\
  \subfloat[]{{\includegraphics[width=2.6in]{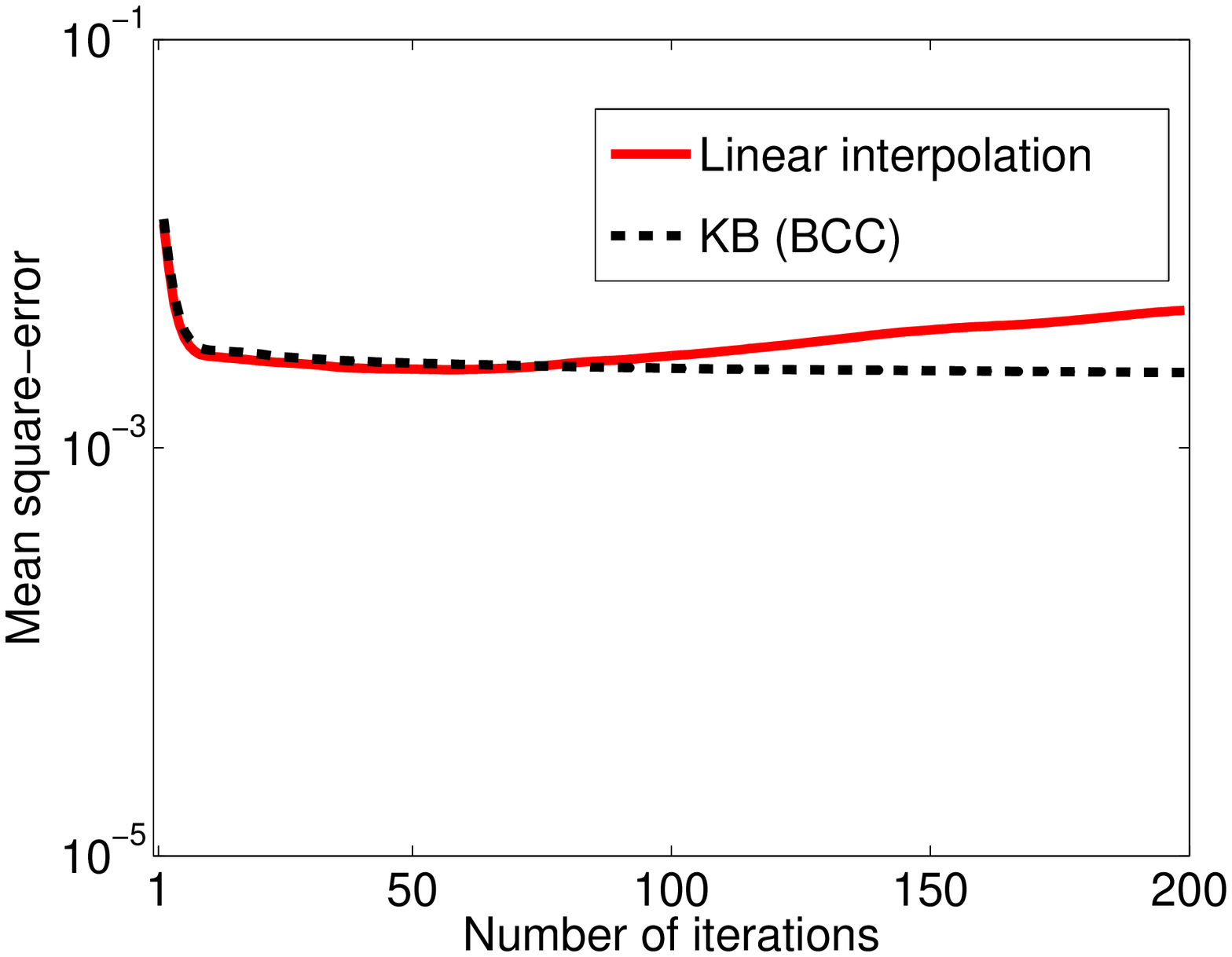}}}\hskip .5cm
  \subfloat[]{{\includegraphics[width=2.5in]{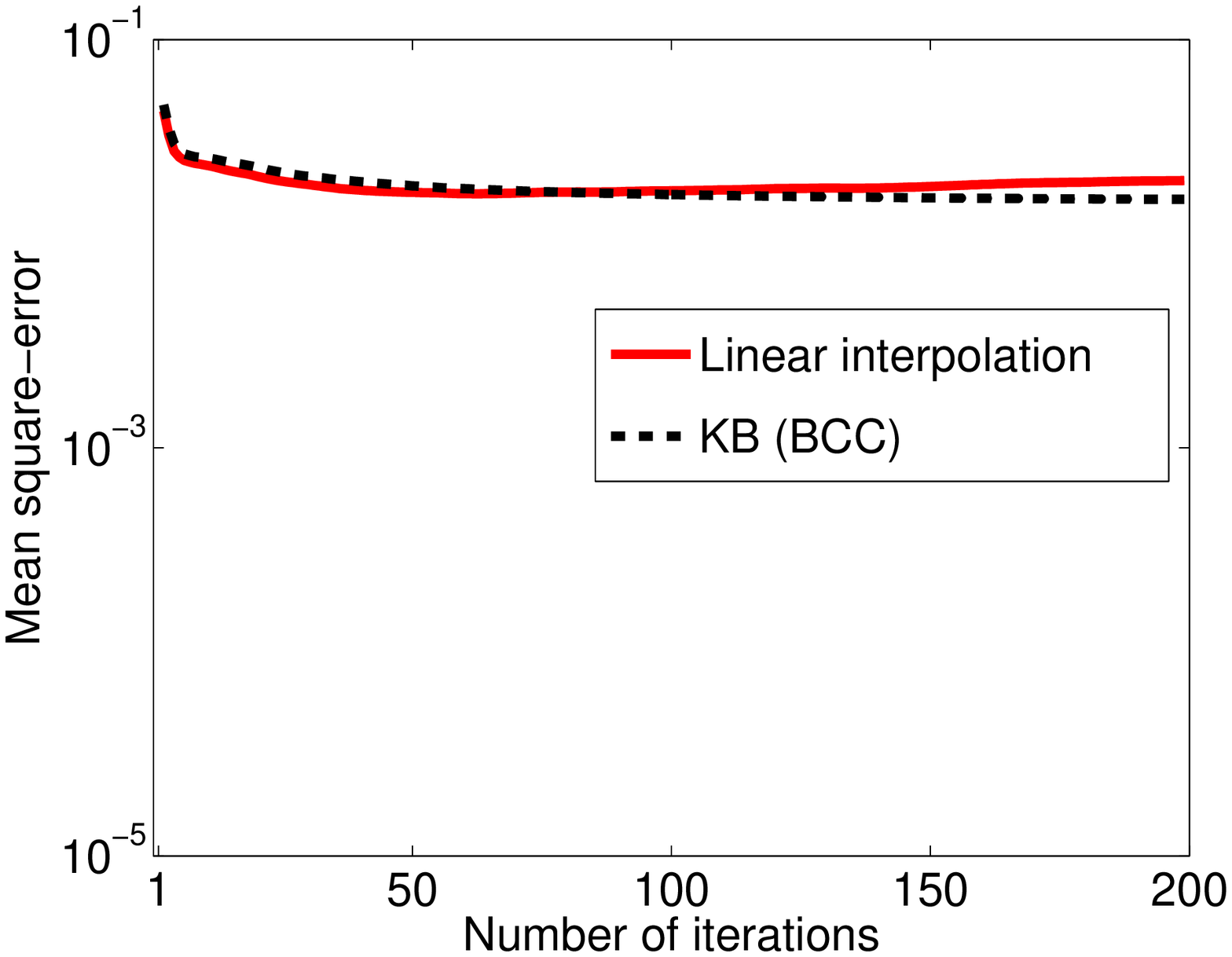}}}
\caption{\label{fig:mse_vs_niter}
Plots of regional MSEs of reconstructed images against the number of iterations, where the MSEs were calculated for 
(a) the plane $z=0$,
(b) the uniform ROI 
(c) the slowly varying ROI
(d) the moderately blurred ROI
(e) the sharp-edge ROI
and 
(f) the sharp, small structure ROI. }
\end{figure}
\clearpage

\begin{figure}[h]
\centering
  \subfloat[]{{\includegraphics[width=2.0in]{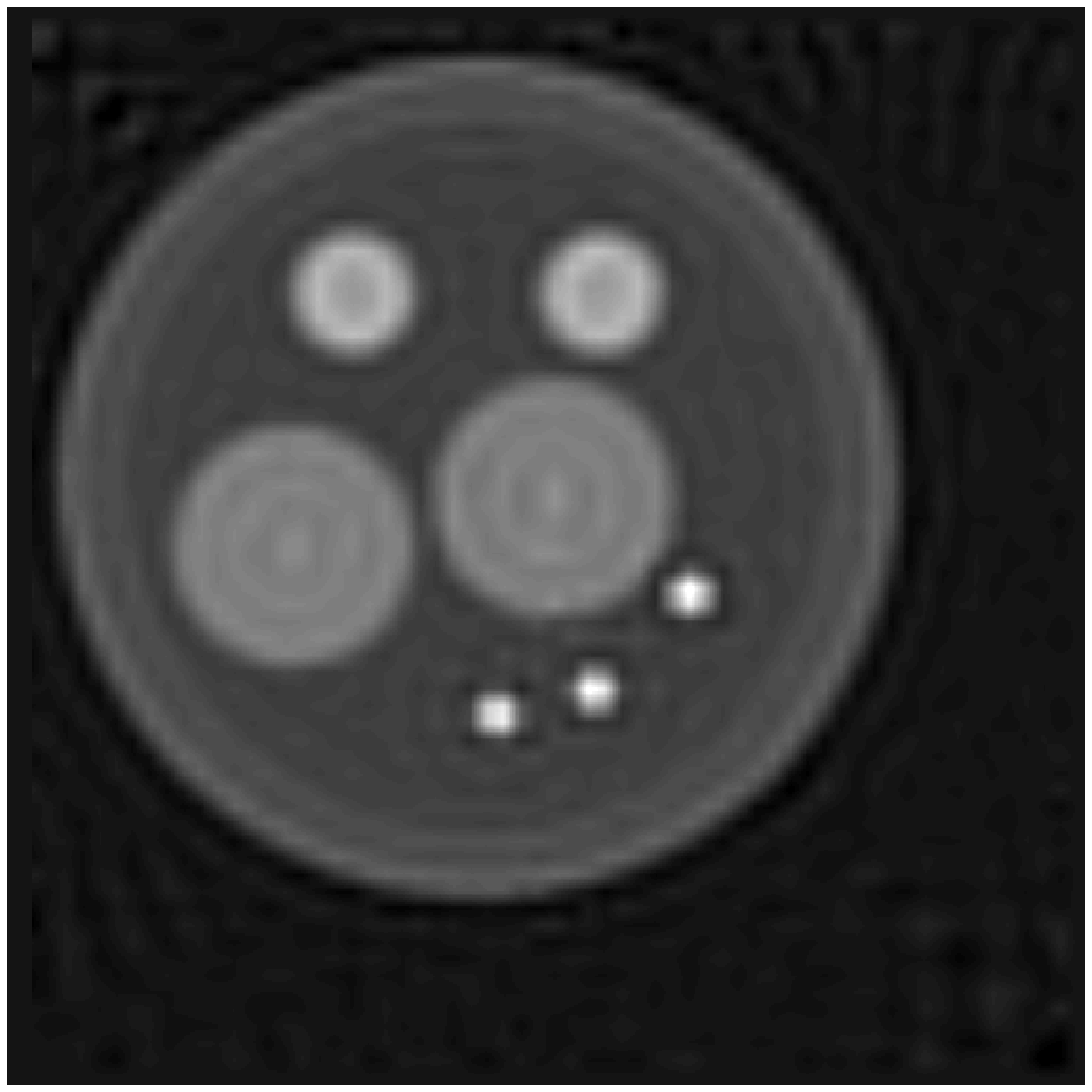}}}\hskip 0.5cm 
  \subfloat[]{{\includegraphics[width=2.0in]{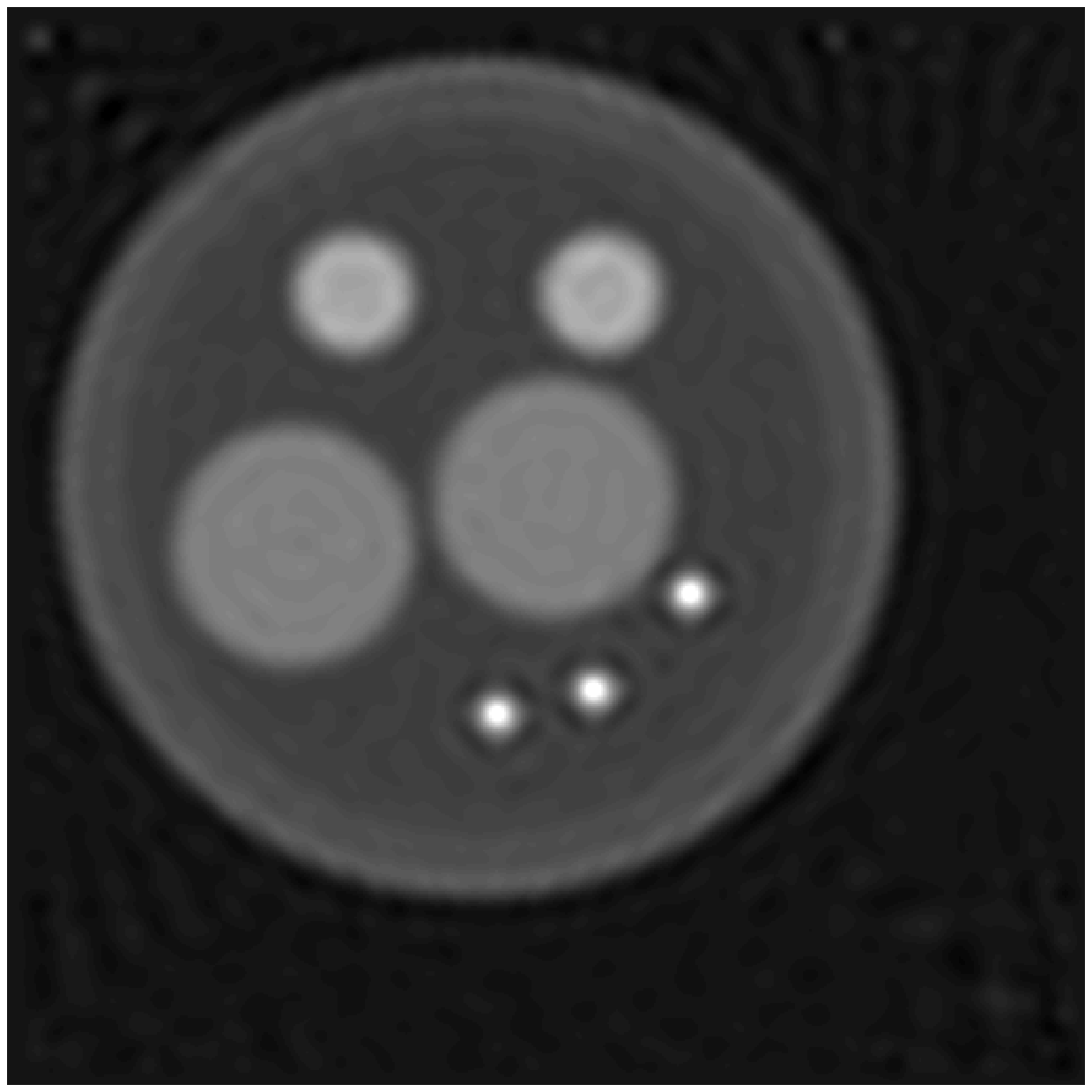}}} \\
  \subfloat[]{{\includegraphics[width=2.5in]{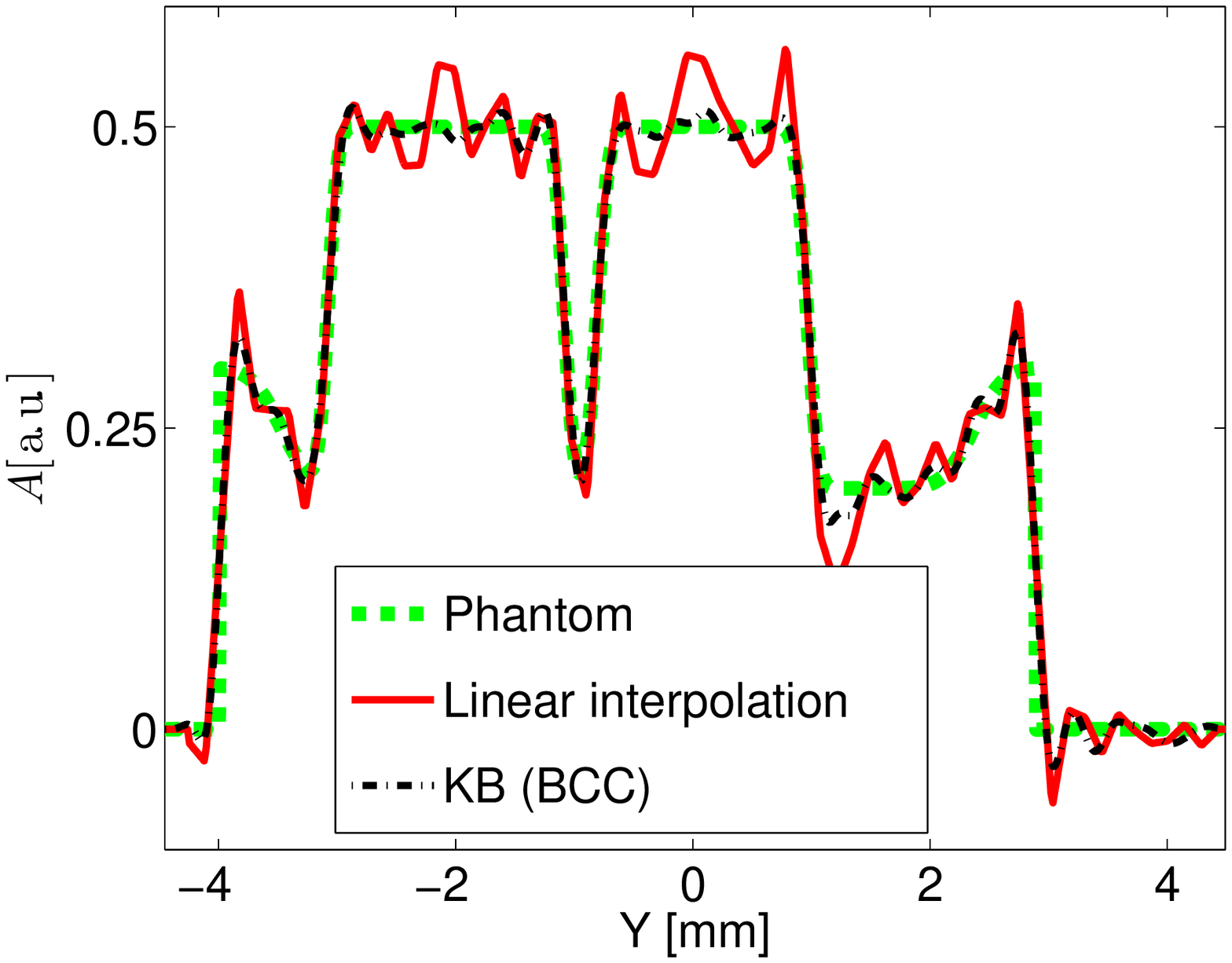}}} \hskip 0.5cm
  \subfloat[]{{\includegraphics[width=2.5in]{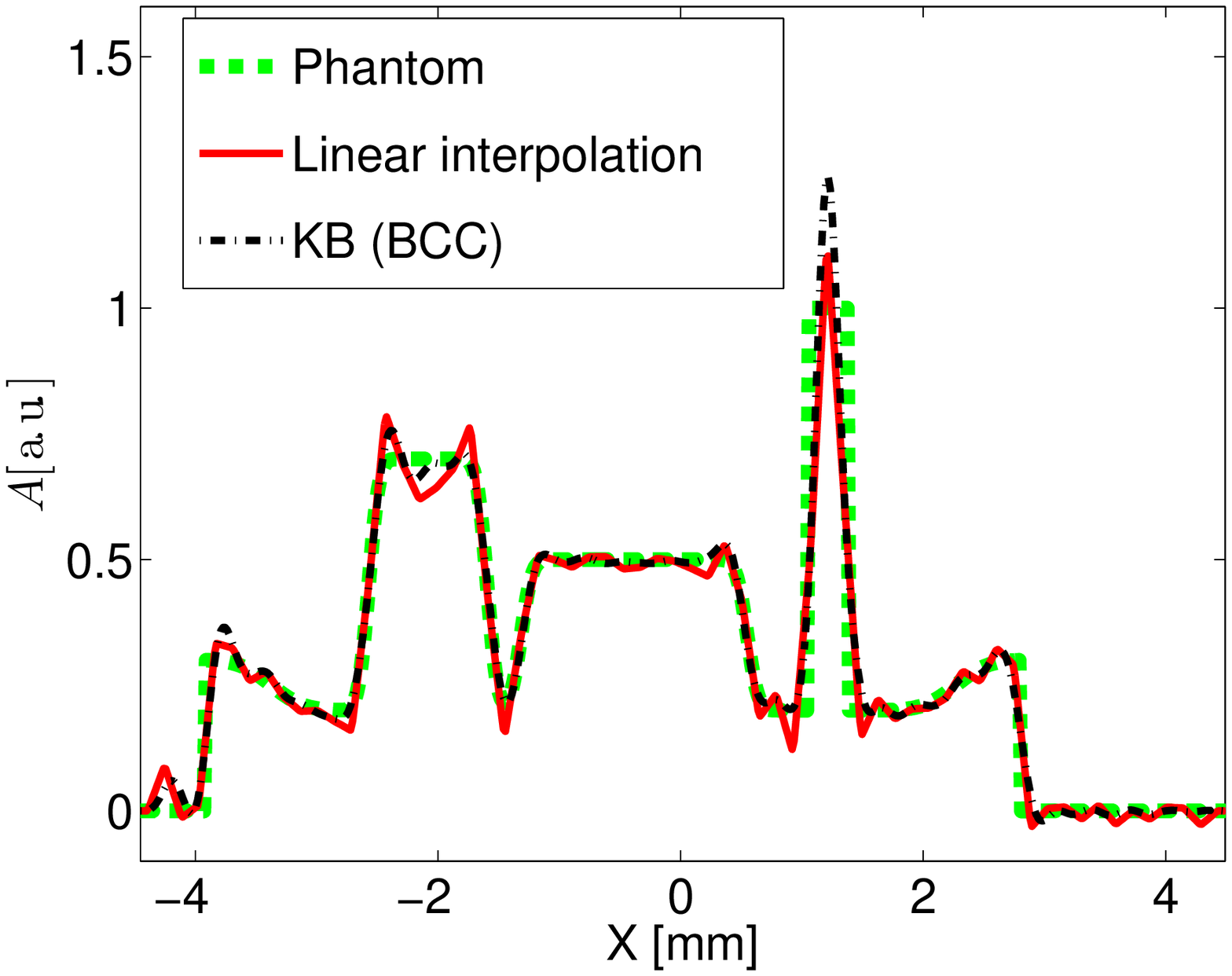}}}
\caption{\label{fig:optmnoiseless}
Slices corresponding to the plane $z=0$ through the $3$D images $\hat{A}_d^{(0)}(\mathbf r)$ reconstructed by use of system matrices 
(a) $\mathbf H_{\rm int}$ and
(b) $\tilde{\mathbf H}_{\rm KB}$ from noise-free data. 
The grayscale window is $[-0.1,1.1]$. 
Profiles of the phantom and reconstructed images are shown along the lines 
(c) $x=0.0788$ mm
and 
(d)  $y=0.429$ mm
in the plane $z=0$.  
Each reconstruction was terminated when the minimal MSE in the plane $z=0$ was achieved. 
}
\end{figure}
\clearpage

\begin{figure}[h]
\centering
  \subfloat[]{{\includegraphics[width=2in]{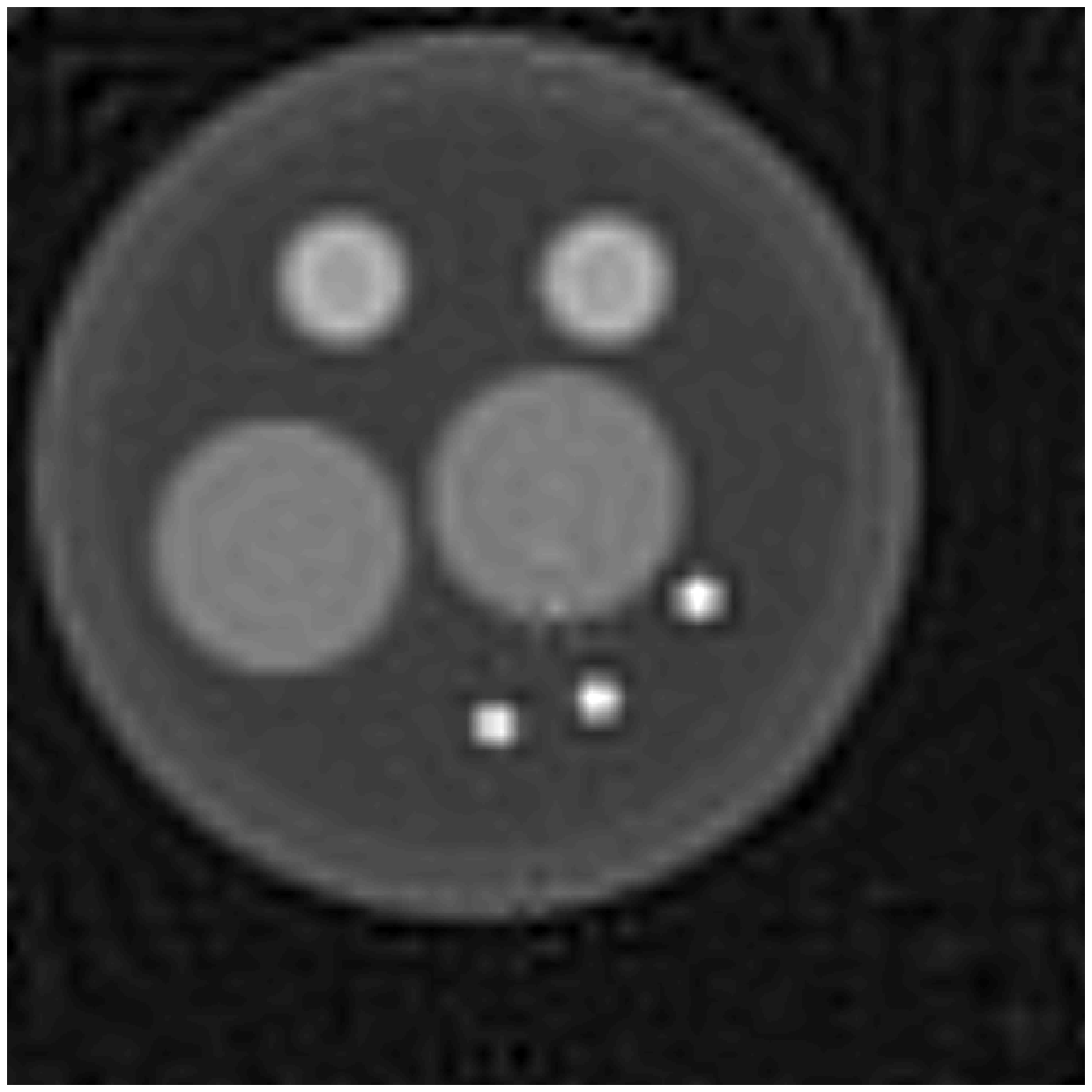}}}
  \subfloat[]{{\includegraphics[width=2in]{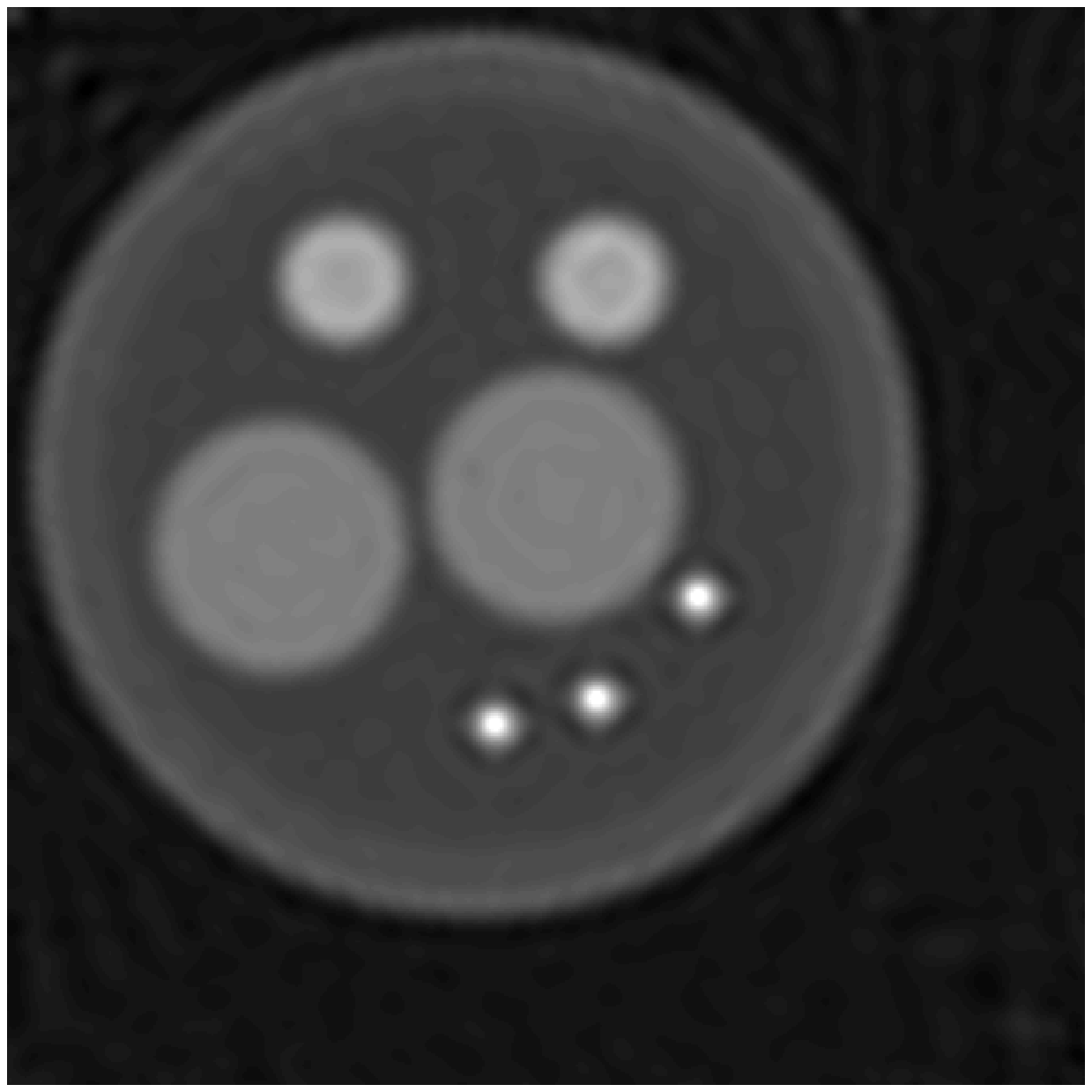}}}\hskip -0.1cm
  \includegraphics[height=2 in]{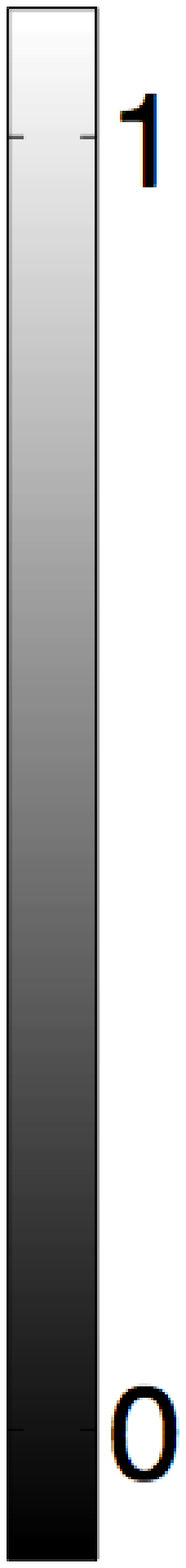}\\
   \hskip 0.6cm
  \subfloat[]{{\includegraphics[width=2in]{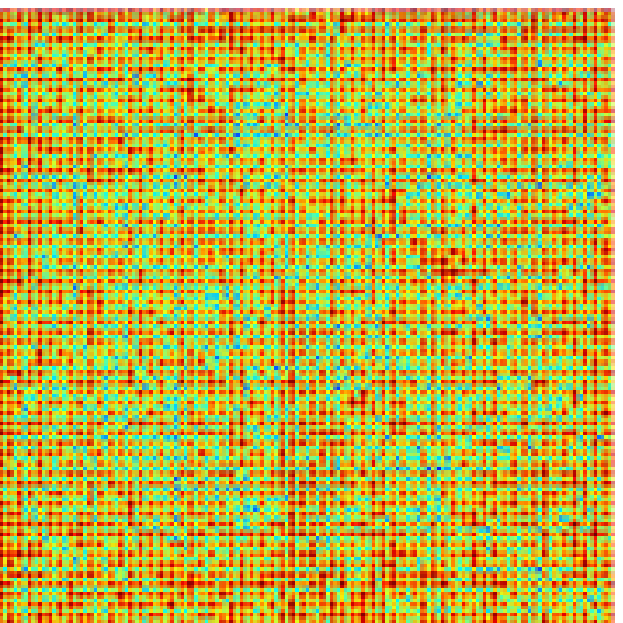}}}
  \subfloat[]{{\includegraphics[width=2in]{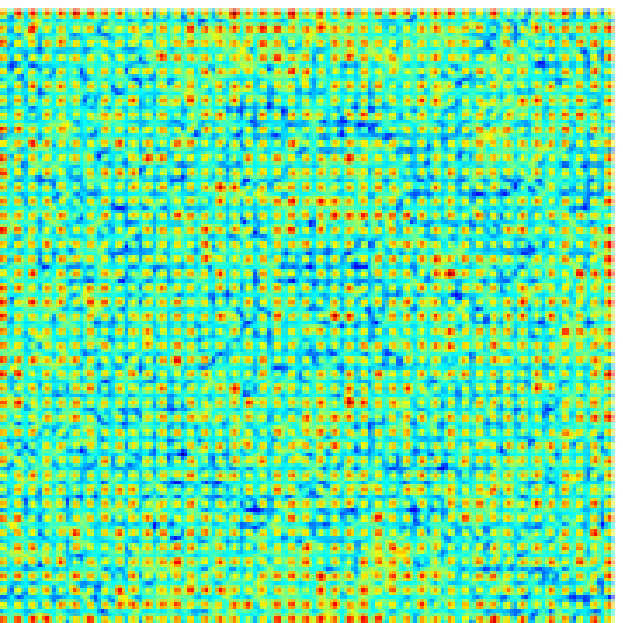}}}\hskip -0.0001 cm
  \includegraphics[height=2 in]{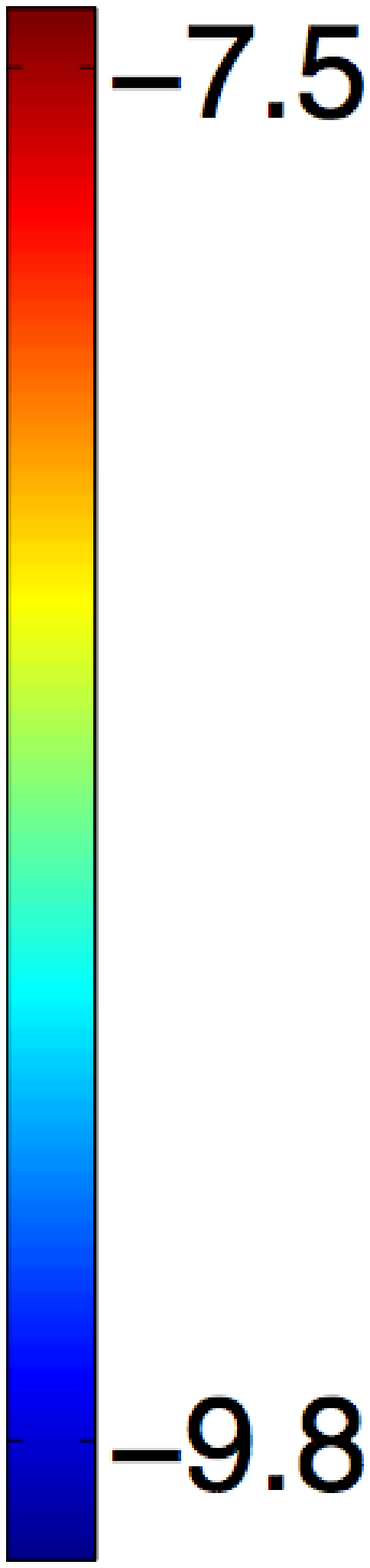}
\caption{\label{fig:noisy_example}
Slices corresponding to the plane $z=0$ through the mean (a-b) and the variance (c-d) of 3D images reconstructed from noisy data, where (a) and (c) correspond to $\mathbf H_{\rm int}$ with $\beta_{\rm int} = 40.0$, while (b) and (d) correspond to $\tilde{\mathbf H}_{\rm KB}$ with $\beta_{\rm KB} = 100.0$. 
The grayscale window for (a) and (b) is $[-0.1, 1.1]$ while the display window for (c) and (d) is on a logarithmic scale ranging from $-10$ (blue) to $-7.4$ (red). 
}
\end{figure}
\clearpage

\begin{figure}[h]
\centering
  \subfloat[]{{\includegraphics[width=3in]{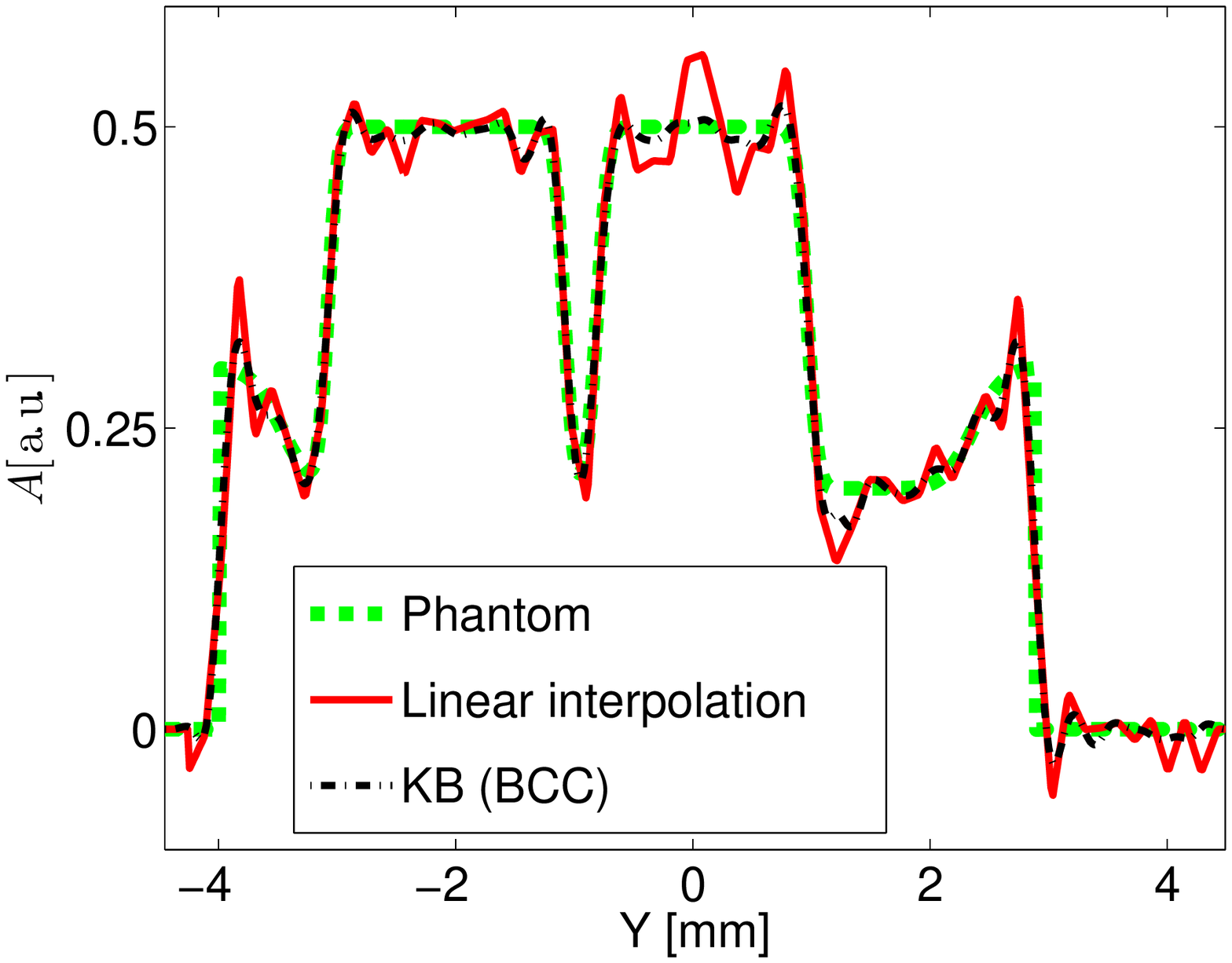}}}
  \subfloat[]{{\includegraphics[width=3in]{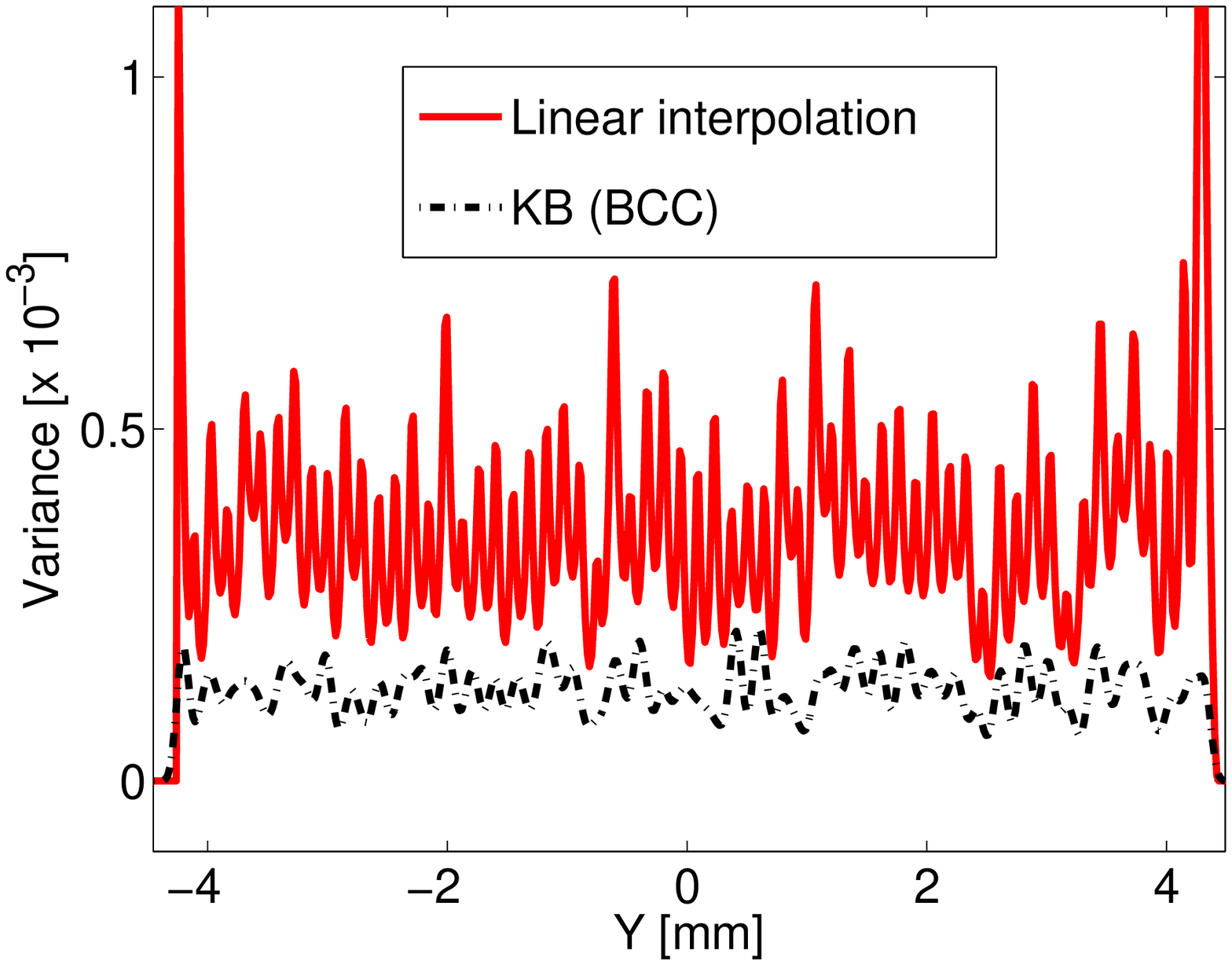}}}
\caption{\label{fig:noisy_profile}
Profiles of (a) the mean and (b) the variance images at $x=0.0788$ mm in the plane $z=0$ from images reconstructed from noisy data using $\mathbf H_{\rm int}$ (red solid line) and $\tilde{\mathbf H}_{\rm KB}$ (black dashed line). 
}
\end{figure}
\clearpage

\begin{figure}[h]
\centering
  \subfloat[]{{\includegraphics[width=2.5in]{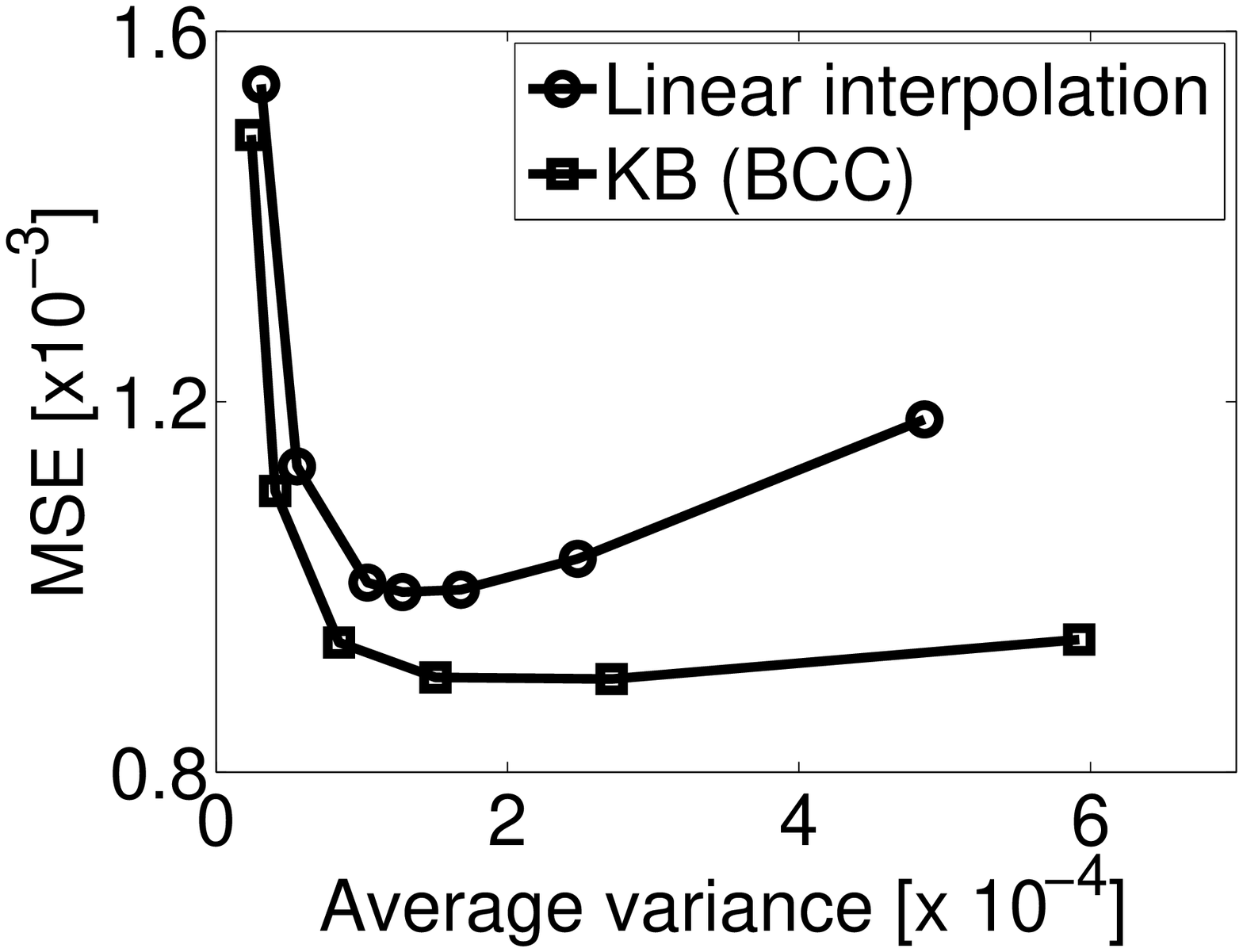}}}\hskip 0.5cm
  \subfloat[]{{\includegraphics[width=2.5in]{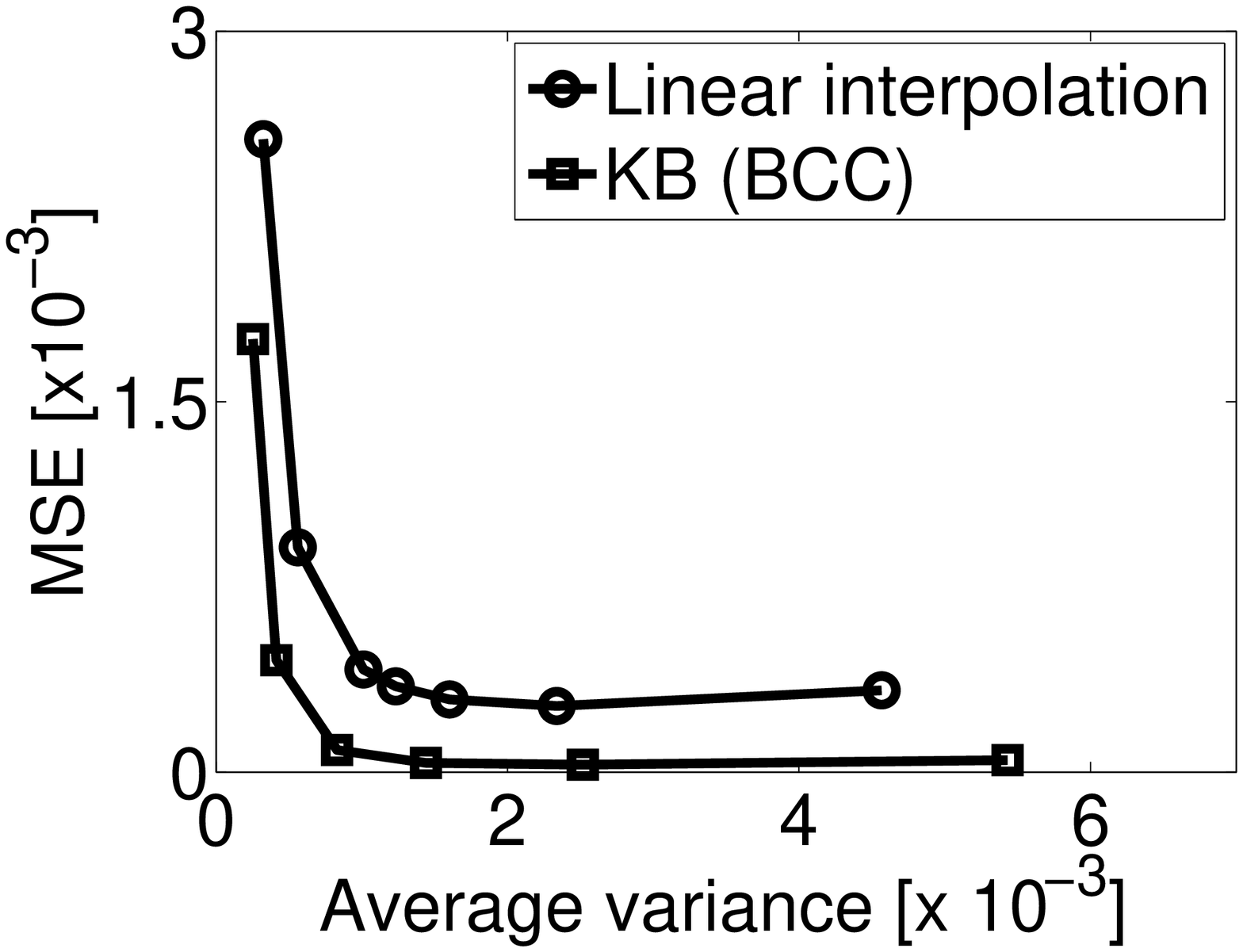}}}\\
  \subfloat[]{{\includegraphics[width=2.5in]{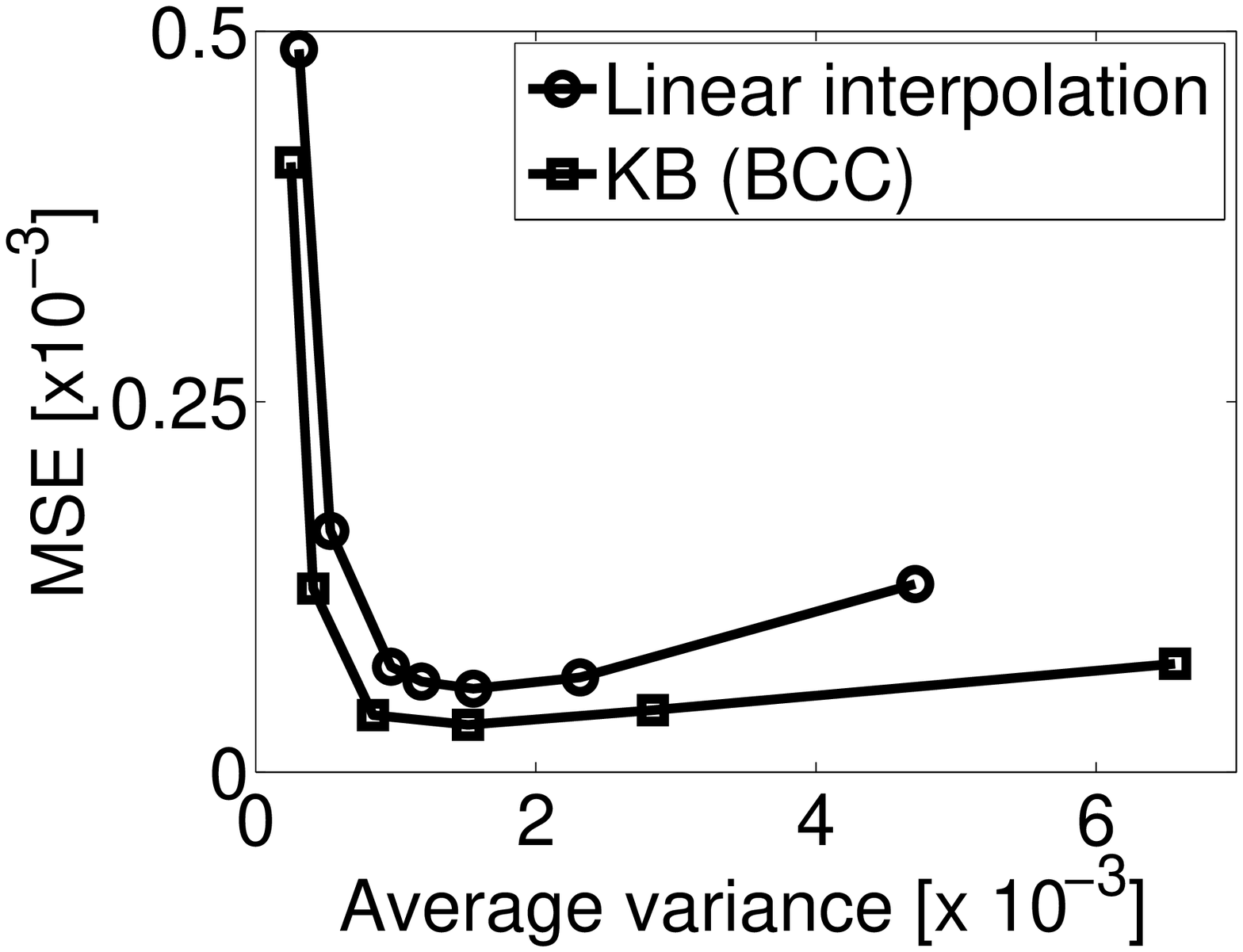}}} \hskip .5cm
  \subfloat[]{{\includegraphics[width=2.5in]{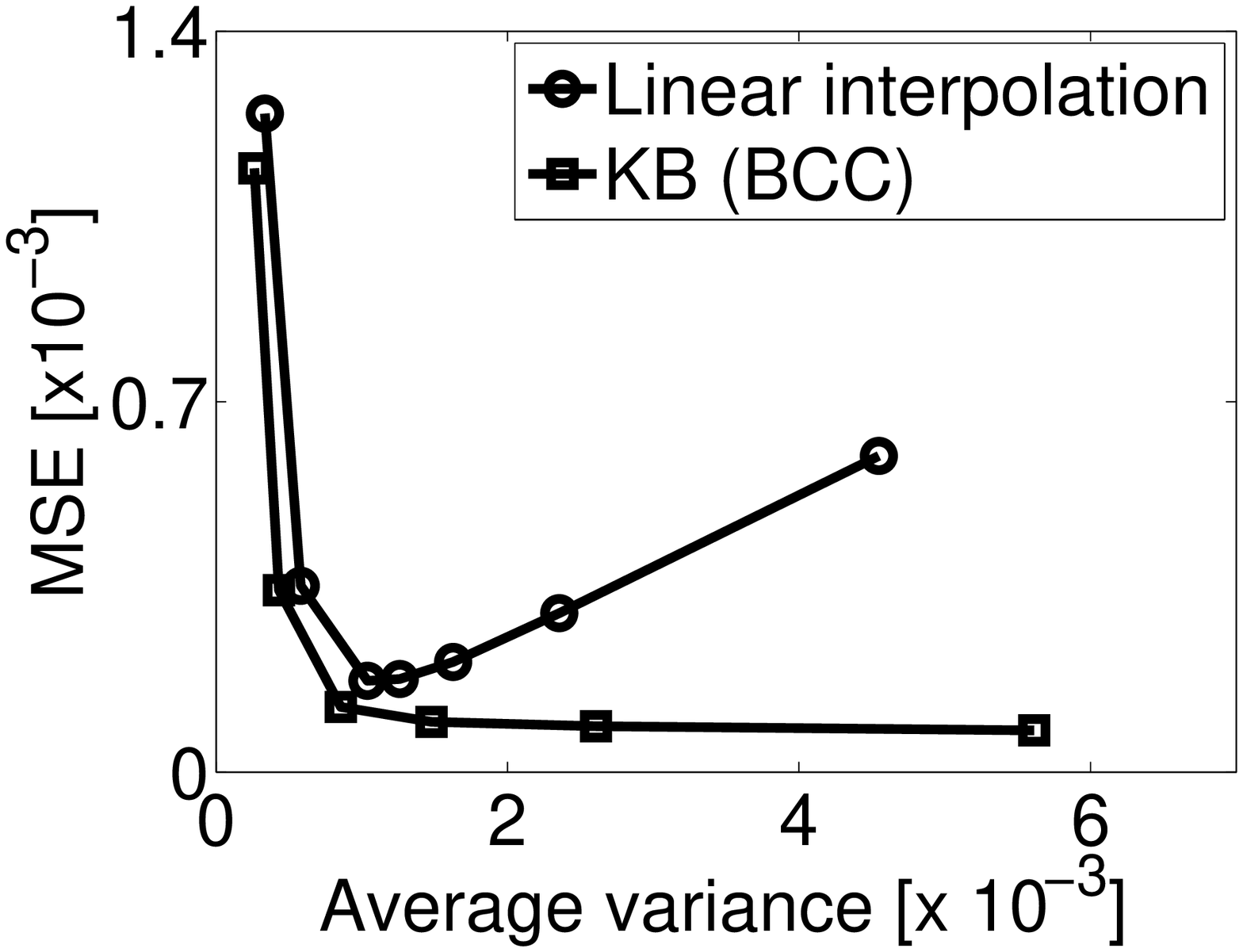}}}\\
  \subfloat[]{{\includegraphics[width=2.5in]{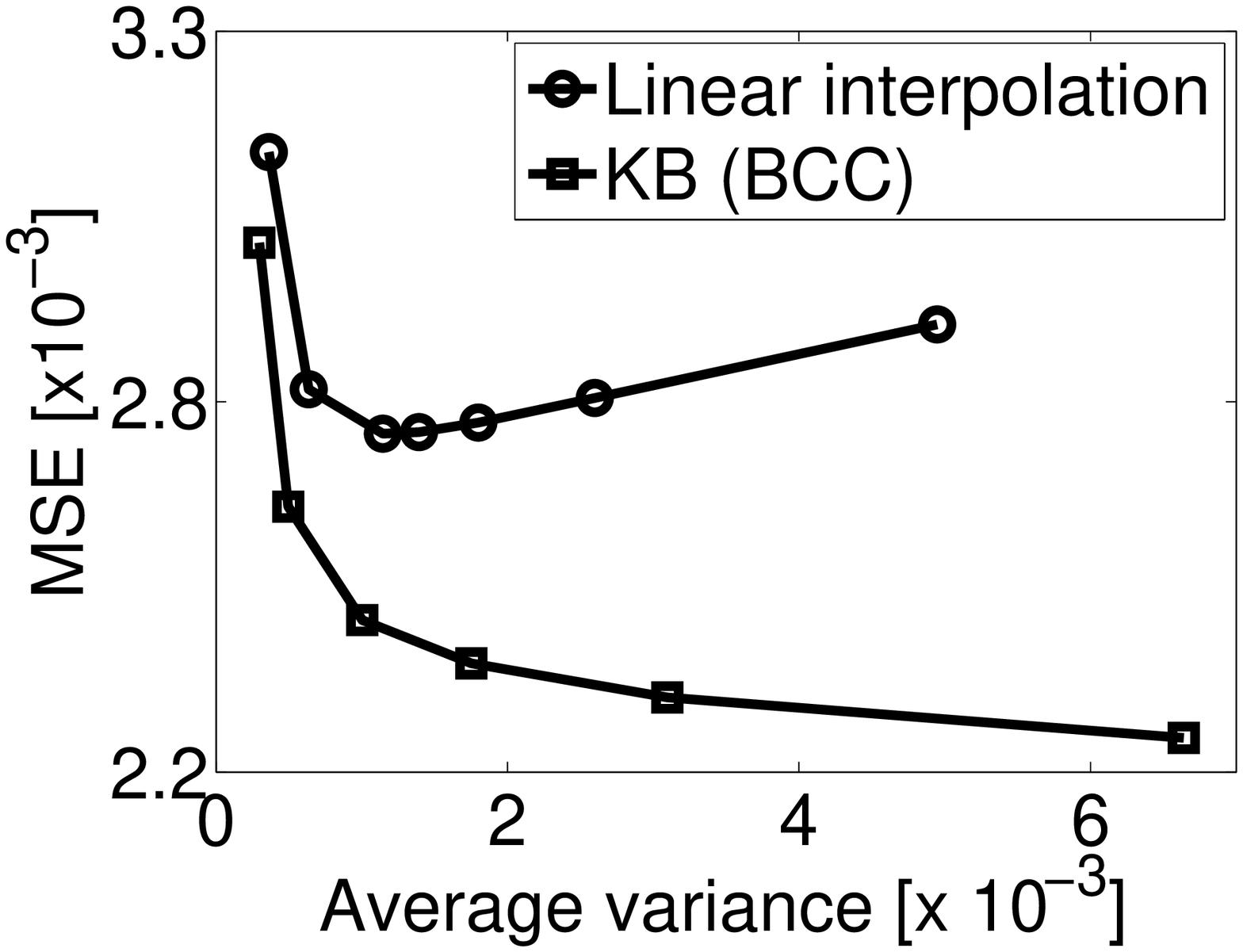}}}\hskip .5cm
  \subfloat[]{{\includegraphics[width=2.5in]{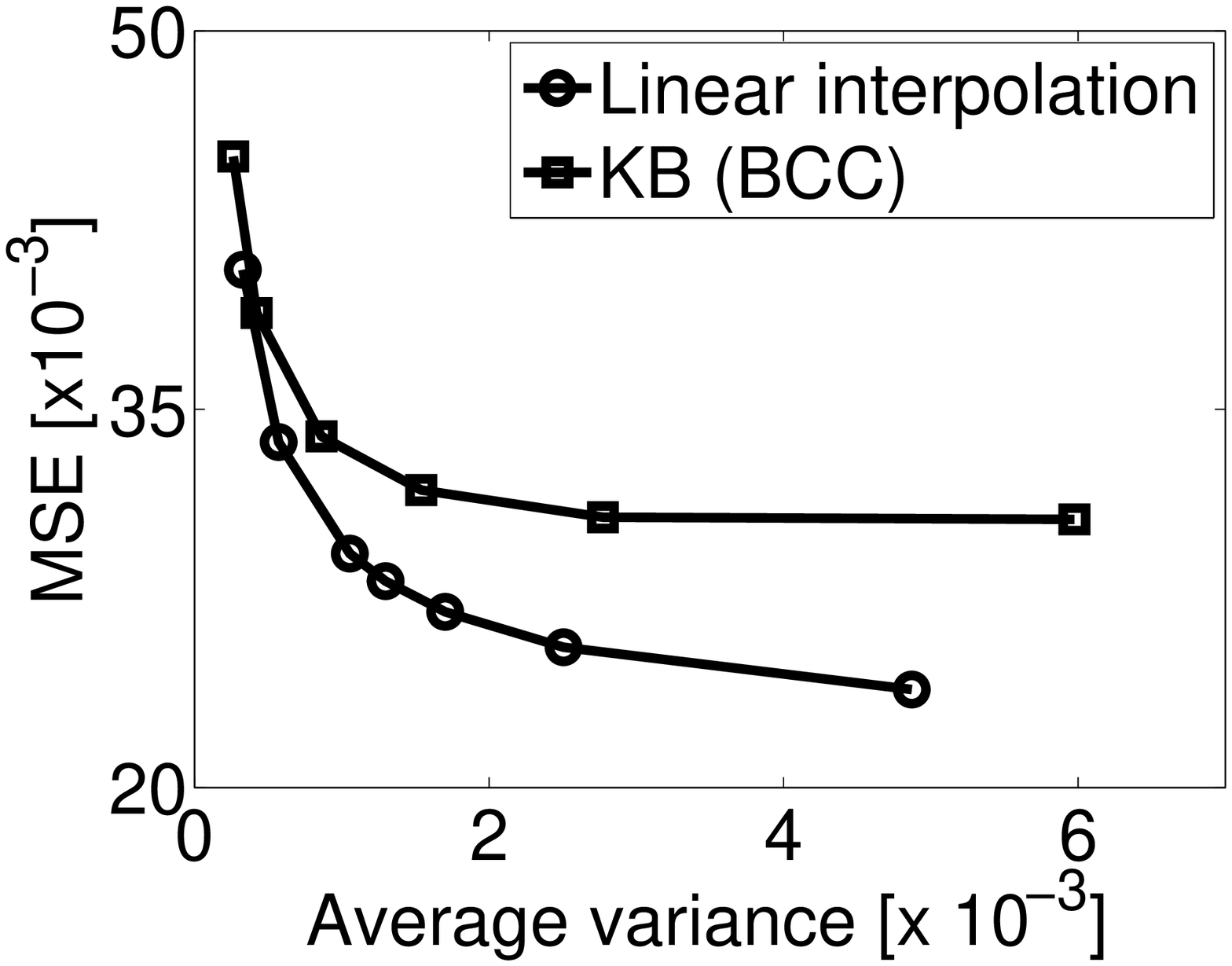}}}
\caption{\label{fig:mse_vs_var}
Plots of the MSEs of mean images against average variances within various regions: 
(a) the plane $z=0$,
(b) the uniform ROI
(c) the slowly varying ROI
(d) the moderately blurred ROI
(e) the sharp-edge ROI
and
(f) the sharp, small structure ROI. }
\end{figure}
\clearpage

\begin{figure}[h]
\centering
  \subfloat[]{{\includegraphics[height=2.6in]{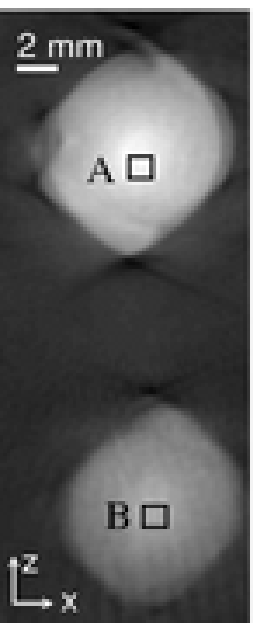}}}\hskip 0.5 cm
  \subfloat[]{{\includegraphics[height=2.6in]{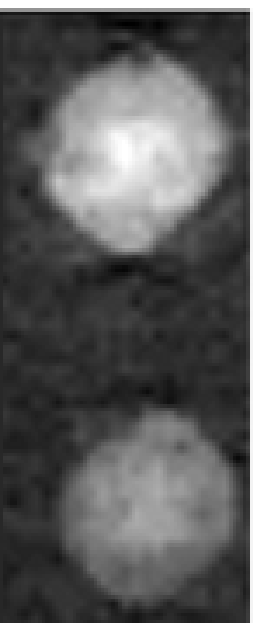}}}\hskip 0.5cm
  \subfloat[]{{\includegraphics[height=2.6in]{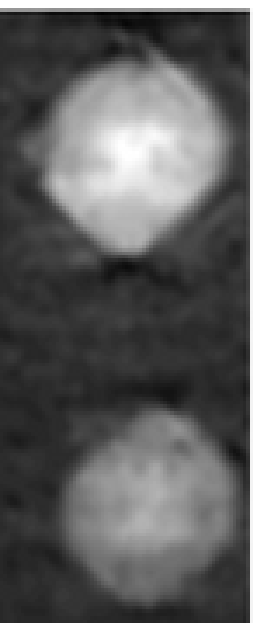}}}
  \includegraphics[height=2.6 in]{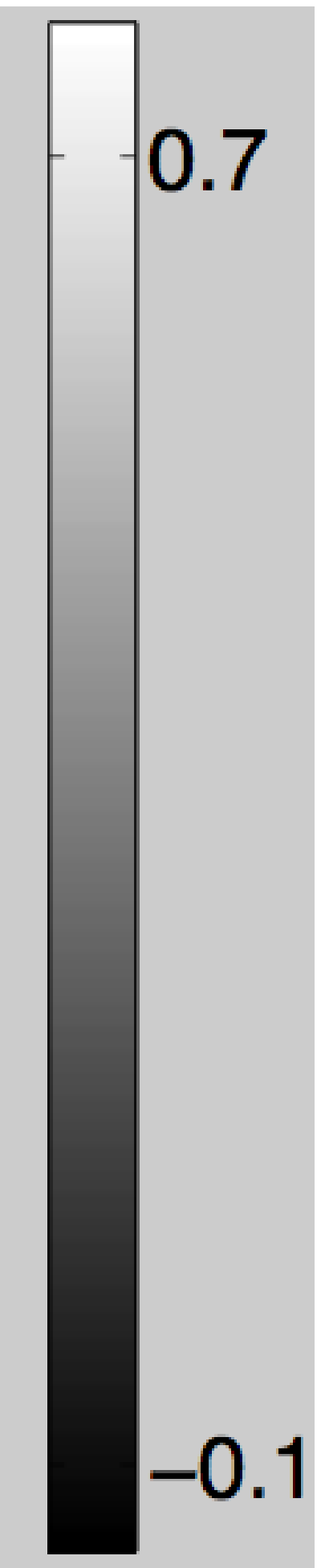}
\caption{\label{fig:exp_single_realization}
Slices corresponding to the plane  $y=1.4$ mm through 
(a) the 3D reference image 
and the 3D images reconstructed by use of 
(b) $\mathbf H_{\rm int}$
and 
(c) $\tilde{\mathbf H}_{\rm KB}$ 
from a single  noisy laboratory measurement.  
The grayscale window is $[-0.16,0.78]$. 
In panel (a), two black boxes mark the ROIs used to conduct the parameter estimation task.}
\end{figure}
\clearpage

\begin{figure}[h]
\centering
  \subfloat[]{{\includegraphics[width=1.2in]{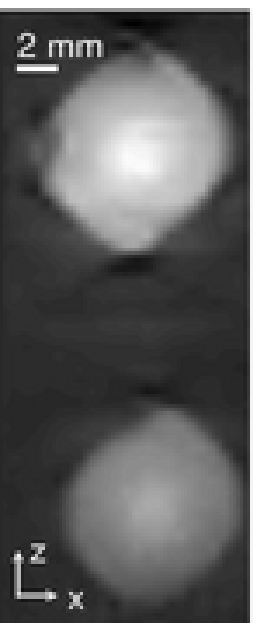}}}\hskip 0.3 cm
  \subfloat[]{{\includegraphics[width=1.2in]{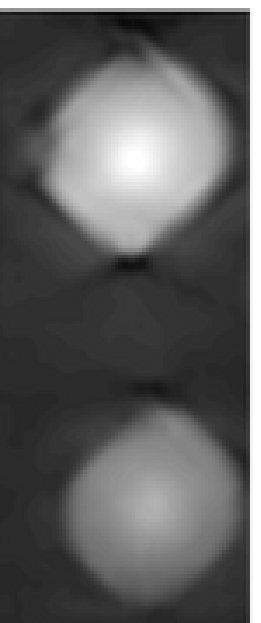}}}\hskip 0.3cm
  \subfloat[]{{\includegraphics[width=1.2in]{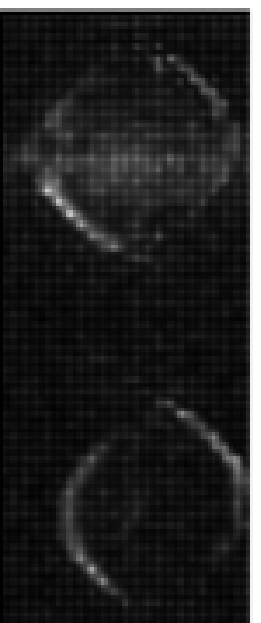}}}\hskip 0.3 cm
  \subfloat[]{{\includegraphics[width=1.2in]{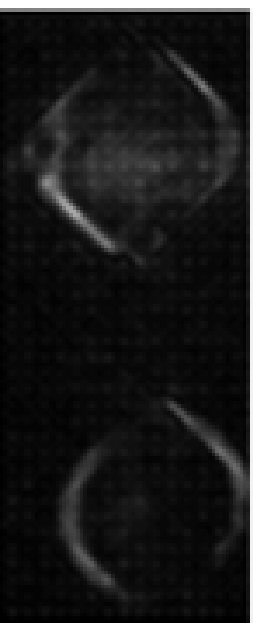}}}\hskip 0.3cm
\caption{\label{fig:exp_mean_var}
Slices through the mean (a-b) and variance (c-d) images corresponding to the plane  $y=1.4$ mm of the 3D images reconstructed from $64$ laboratory measurements. 
The mean images correspond to 
(a) $\mathbf H_{\rm int}$ with $\Delta_{\rm s}=0.56$ mm, 
and 
(b) $\tilde{\mathbf H}_{\rm KB}$ with $\Delta{\rm ^b_s}=0.8$ mm 
and use the same grayscale window of $[-0.16,0.78]$. 
The variance images found when using (c) $\mathbf H_{\rm int}$ and (d)  $\tilde{\mathbf H}_{\rm KB}$ use the grayscale window of $[0,8.0\times10^{-3}]$. 
}
\end{figure}
\clearpage

\begin{figure}[h]
\centering
  \subfloat[]{{\includegraphics[width=3in]{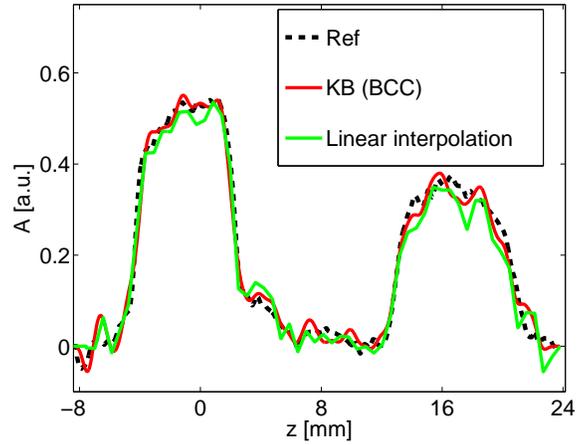}}}\\
  \subfloat[]{{\includegraphics[width=3in]{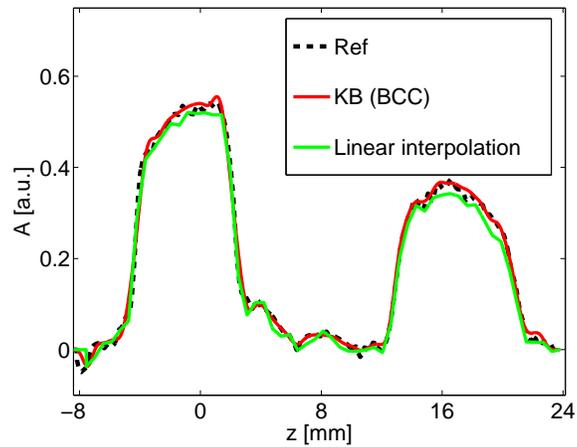}}}
\caption{\label{fig:exp_profile}
Profiles of 3D images along the line $y=1.4$ mm, $x=2.1$ mm for the reference (black dashed line) and reconstructions using $\tilde{\mathbf H}_{\rm KB}$ (solid red line) and $\mathbf H_{\rm int}$ (solid green line).   Profiles are shown for images reconstructed from (a) a single laboratory measurement
and (b) the mean image averaged over $64$ laboratory measurements.  
}
\end{figure} 
\clearpage

\begin{figure}[h]
\centering
  \subfloat[]{{\includegraphics[width=3in]{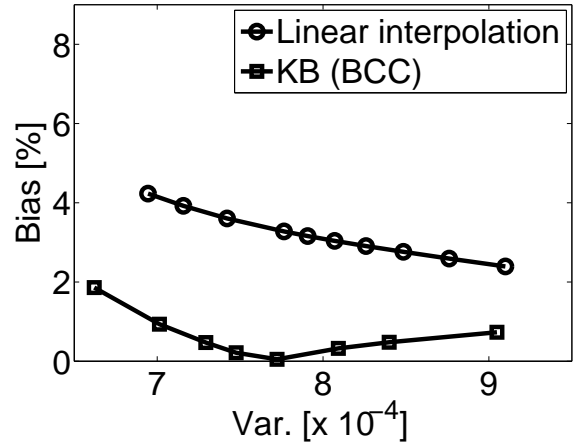}}}\\
  \subfloat[]{{\includegraphics[width=3in]{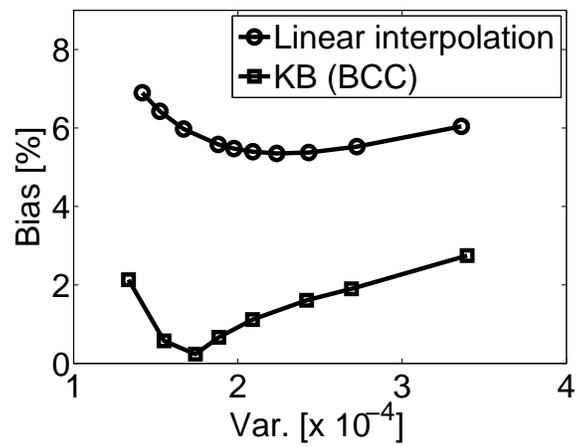}}}
\caption{\label{fig:exp_biasvsvar}
Plots of bias against variance from the experimental data sets within (a) ROI-A and (b) ROI-B as defined in Fig. ~\ref{fig:exp_single_realization}.
}
\end{figure} 

\end{document}